\documentclass[aps,prb,showpacs,floatfix,amsmath,amssymb,superscriptaddress,reprint]{revtex4-2}
\usepackage{bbold}
\usepackage[normalem]{ulem}
\usepackage{color}
\usepackage{graphicx}
\usepackage{epstopdf}
\usepackage{setspace}
\usepackage{amsmath}
\usepackage{amssymb}
\usepackage{verbatim}
\usepackage{cancel}
\usepackage{xcolor}

\definecolor{scarred}{rgb}{0.75,0.0,0.0}
\usepackage{enumitem} 
\usepackage[colorlinks=true,citecolor=blue,linkcolor=blue,urlcolor=blue]{hyperref}
\usepackage{seqsplit}

\begin{document}
\title{Quantitative DFT+DMFT description of spectra and transport in the moderately correlated metal SrVO$_3$}
\author{Gurshidali\ P.}
\email{gurshidali@gmail.com}
\author{N.\ S.\ Vidhyadhiraja}
\email{raja@jncasr.ac.in}
\affiliation{Theoretical Science Unit, Jawaharlal Nehru Centre for Advanced Scientific Research, Bangalore, 560064, India}

\begin{abstract}
A quantitative, material-specific account of spectral and transport properties remains a central challenge in the theory of strongly correlated materials. Combining density functional theory with dynamical mean-field theory (DFT+DMFT) has proven to be a powerful approach for treating electron correlation effects and material specificity on an equal footing. Here, we examine the single-particle spectra and the dc and optical conductivity of SrVO$_3$, a prototypical, moderately correlated metal, within this framework. The degenerate $t_{2g}$ active space of SrVO$_3$, together with its well-established Fermi-liquid behavior, admits an effective-mass description governed by a single quasiparticle weight, yielding a nearly universal picture of the low-frequency, low-temperature regime. Employing a computationally efficient, real-frequency multi-orbital iterative perturbation theory (MO-IPT) impurity solver, we find reasonable agreement with experimental measurements of dc resistivity and optical conductivity across the entire experimentally relevant $(\omega, T)$ range within a single, unified scheme. The agreement is shown to not depend on specific interaction parameters provided the quasiparticle weight is kept constant. These results indicate that, in SrVO$_3$, the $\mathbf{k}$-dependence of the self-energy may be weak, and vertex corrections may not dominate the dc and optical transport in this material.
\end{abstract}
\maketitle

\section{Introduction}
\label{sec:intro}
Strongly correlated electron systems exhibit a rich interplay between electronic structure and many-body interactions, giving rise to electronic properties that cannot be understood within conventional band theory alone~\cite{Georges_RMP1996, Imada_RMP1998}. Over the past two decades, the combination of density functional theory with dynamical mean-field theory (DFT+DMFT) has emerged as a powerful framework for incorporating material-specific electronic structure and local electronic correlations on an equal footing~\cite{Kotliar_RMP2006,Held2007,Basov_RMP2011}. While this approach has been remarkably successful in describing single-particle spectral properties, achieving a quantitative and unified description of both spectral and transport properties in realistic multiorbital materials over all experimentally relevant frequency and temperature scales remains a significant challenge. Addressing this problem forms the central motivation of the present work.

Among correlated materials, the cubic perovskite SrVO$_3$ has emerged as a prototypical moderately correlated metal and a model system for realistic many-body calculations~\cite{Fujimori_PRL1992,Nekrasov_PRB2005,Taranto_PRB2013}. Its simple electronic structure, well-isolated $t_{2g}$ manifold, and extensive experimental characterization make it an ideal platform for quantitative comparisons between theory and experiment~\cite{Sekiyama_PRL2004,Aizaki_PRL2012,Brahlek2024}.
Beyond its role as a benchmark correlated metal, SrVO$_3$ has recently attracted considerable interest as a transparent conducting oxide, combining metallic conductivity with optical transparency~\cite{APL2018,Xu2019,PRM2021, ACS2023, ACS2024}.

On the theoretical side, DFT+DMFT has achieved considerable success in describing the single-particle spectral properties of SrVO$_3$, capturing mass renormalization, bandwidth reduction, and Hubbard bands~\cite{Fujimori_PRB1995,Inoue_PRB1998,Yoshida_PRL2005,Nekrasov_PRB2005,Aizaki_PRL2012}. Transport properties have been investigated using a range of approaches, including Boltzmann theory~\cite{Smirnov2022,BoltzTransport2003,BoltzTrap2006}, Kubo-formalism-based calculations~\cite{Assmann2016,GAhn_PRB2022,Brahlek2024,Abramovitch2024,Kugler_PRB2026,LaBolita2026}, and approximate treatments of scattering mechanisms~\cite{Inoue_PRB1998,Cvijovic2011,AshcroftMermin,Ziman1960}. These studies are typically carried out within distinct theoretical frameworks and often address spectral and transport properties separately. Most material-specific DFT+DMFT studies of SrVO$_3$ employ continuous-time quantum Monte Carlo (CTQMC) impurity solvers~\cite{Werner_PRL2006, EGull_RMP2011}. Although numerically exact and unbiased in principle, these solvers are subject to inherent challenges, such as the necessity of analytical continuation, the fermionic sign problem, and exponential scaling. For multi-orbital correlated systems, these computational constraints significantly hinder direct comparisons between theory and experiment, particularly in the low-temperature regime ($T\lesssim 100K$).

Alternatively, the numerical renormalization group offers a deterministic, 
real-frequency framework that avoids analytical continuation and the sign problem 
while providing high resolution at low energies via logarithmic bath 
discretization~\cite{Wilson1975,NRG_RMP2008}. However, its application to 
multiorbital systems is constrained by exponential Hilbert 
space scaling.More recently, the ghost rotationally invariant slave-boson (gRISB) method~\cite{gRISB2023,gRISB2024} has emerged as an efficient auxiliary-particle approach for solving multi-orbital Hubbard models. Although it accurately captures low-energy quasiparticle renormalization, its mean-field formulation does not explicitly describe the full frequency dependence of the electronic self-energy. Consequently, phenomena associated with dynamical scattering and incoherent spectral weight redistribution are not treated on the same footing as in fully dynamical impurity solvers, limiting its applicability to finite-frequency spectroscopic and transport properties such as the optical conductivity and dc resistivity~\cite{slaveboson1986,Rohringer_RMP2018,slaveboson2020}.

In light of these challenges, a clear methodological gap persists between the experimentally well-characterized spectral and transport properties of SrVO$_3$ and their theoretical description within contemporary many-body frameworks. The multi-orbital iterated perturbation theory (MO-IPT) solver employed in this work is based on an analytical ansatz that has been extensively benchmarked against numerically exact CTQMC calculations~\cite{Dasari_EPJB2016}. Previous studies have demonstrated good agreement for a broad class of correlated systems, including single- and multi-orbital Hubbard models and the material-specific $t_{2g}$ manifold of SrVO$_3$, where the calculated spectral properties compare favorably with both CTQMC and photoemission experiments~\cite{Dasari_EPJB2016}. As with any approximate impurity solver, its accuracy depends on the underlying parameter regime. For example, the agreement with CTQMC improves as one moves away from the particle-hole symmetric limit of the degenerate multi-orbital Hubbard model~\cite{Dasari_EPJB2016}.

For the present study of SrVO$_3$, since the degenerate $t_{2g}$ manifold is at one-third filling~\cite{Fujimori_PRL1992,Fujimori_PRB1995}, and hence far from p-h symmetry, MO-IPT provides an attractive balance between accuracy and computational efficiency. Embedded within the DFT+DMFT framework~\cite{MRodero_PRB1986,Kajueter_PRL1996,Fujiwara_PRB2008,Dasari_EPJB2016}, it yields deterministic real-frequency self-energies directly, thereby avoiding analytic continuation. Its polynomial computational scaling enables a systematic exploration of the $(U,J)$ parameter space while providing reliable access to the low-temperature regime, making it particularly well suited for the quantitative investigation of spectral and transport properties undertaken in this work.

In the present work, we investigate whether a single material-specific DFT+DMFT framework can provide a quantitative description of both the spectral and transport properties of SrVO$_3$. Using the MO-IPT impurity solver within DFT+DMFT, we solve the material-specific multi-orbital Hubbard model for SrVO$_3$ over a wide range of interaction parameters and across experimentally relevant frequencies and temperatures. Particular emphasis is placed on determining whether the low-energy electronic structure can be characterized by a single quasiparticle renormalization factor and whether this same coherence scale consistently describes both the optical conductivity and dc transport. We show that an approximate universal scaling emerges across these observables, indicating that a single quasiparticle coherence scale captures much of the low-energy physics of SrVO$_3$.

To assess the reliability of the approach, the calculated optical conductivity and dc resistivity are carefully benchmarked against experiment, while the low-frequency self-energy is compared with both experimentally extracted self-energies and numerically exact continuous-time quantum Monte Carlo (CTQMC) calculations. Overall, good agreement is obtained for both the self-energy and the transport properties across the relevant energy and temperature scales. In addition, we present a detailed microscopic analysis of the optical response, identifying interorbital hybridization within the $t_{2g}$ manifold as the origin of the low-energy interband optical feature. Together, these results demonstrate that MO-IPT, embedded within a material-specific DFT+DMFT framework, provides a quantitative and computationally efficient approach for describing the spectral and transport properties of the moderately correlated metal, SrVO$_3$.

The remainder of this paper is organized as follows. Sec.~\ref{sec:methodology} presents the formalism of the real-frequency multi-orbital Iterative Perturbation Theory (MO-IPT) impurity solver together with the material-specific DFT+DMFT methodology employed for SrVO$_3$. Sec.~\ref{sec:results} presents the results. We first develop the quasiparticle description of the low-energy electronic structure and examine the evolution of the quasiparticle renormalization in the $(U,J)$ parameter space. The calculated self-energies are then benchmarked against both numerically exact CTQMC calculations and experimentally extracted self-energies, followed by a systematic comparison of the calculated dc resistivity and optical conductivity with experiment. The analysis is subsequently extended to the interacting local and momentum-resolved spectral functions, establishing quasiparticle renormalization and adiabatic continuity with the underlying DFT electronic structure. Finally, Section IV summarizes the main conclusions and discusses the broader implications of the present work.

\section{Model and Formalism}
\label{sec:methodology}
We begin by constructing a material-specific low-energy Hamiltonian for SrVO$_3$ using DFT-based Wannierization, which accurately captures the $t_{2g}$ manifold near the Fermi level. The electronic structure of cubic SrVO$_3$ was computed within density functional theory (DFT) using the \textsc{Quantum ESPRESSO} package~\cite{QE2009,QE2017}. Plane-wave basis sets with a kinetic energy cutoff of 60~Ry and ultrasoft pseudopotentials within the generalized gradient approximation (GGA) of Perdew-Burke-Ernzerhof (PBE)~\cite{Perdew_PRL1996} were employed. Brillouin-zone was sampled using a $10\times10\times10$ Monkhorst-Pack $k$-point mesh, and the experimental cubic lattice constant ($a = 3.84$~\AA) was adopted.

To construct a low-energy Hamiltonian, the Kohn-Sham states were projected onto maximally localized Wannier functions for the V-$t_{2g}$ manifold using \textsc{Wannier90}~\cite{Marzari_PRB1997, Souza_PRB2001, Marzari_RMP2012, Mostofi_Elsevier2014}. Disentanglement and localization were performed via spread minimization to accurately reproduce the near-Fermi-level bands. This Wannier representation provides both a physically transparent identification of the correlated subspace and a flexible interface between DFT and dynamical mean-field theory (DMFT), following the formalism introduced by Lechermann \emph{et al.}~\cite{Lechermann_PRB2006}. The resulting Wannier Hamiltonian serves as the non-interacting input for subsequent DMFT calculations.

In SrVO$_3$, the ideal cubic perovskite structure (space group $Pm\bar{3}m$) ensures linear V--O--V bonds ($\sim180^\circ$), maximizing the hybridization between V-$3d$ and O-$2p$ states and producing a broad $t_{2g}$ manifold~\cite{Lan2003}. The resulting Wannier Hamiltonian, obtained from our DFT-based Wannierization, exhibits nearly degenerate on-site energies and a hopping network consistent with the directional character of the $t_{2g}$ orbitals, yielding an overall bandwidth of $W\simeq2.5$~eV. The moderate correlation strength in SrVO$_3$ is reflected in a quasiparticle mass renormalization of $m^*/m\sim2$, as established by photoemission experiments and DFT+DMFT studies~\cite{Fujimori_PRL1992,Fujimori_PRB1995,KMaiti_EPL2001,Sekiyama_PRL2004,Yoshida_PRL2005,Nekrasov_PRB2005,Aizaki_PRL2012,Taranto_PRB2013,Kobayashi2017}. Unlike extremely correlated Kondo insulators and quantum-critical heavy-fermion systems, where emergent many-body energy scales can be exponentially small, and a universal description of spectra and transport has been possible without invoking material-dependent parameters~\cite{Raja_JPCM2003,Raja_EPJB2004,Raja_JPCM2005,Logan2005,Gegenwart2007,Coleman2015}, 
a quantitative description of the moderately correlated SrVO$_3$ would require incorporating material-specific electronic structure and local electronic correlations on equal terms.

To describe the correlated low-energy electronic structure, we consider a multi-orbital Hubbard model with density-density interactions. The Hamiltonian in second quantization is expressed as
\begin{align}
\hat{\mathcal{H}} 
&= \sum_{i,\alpha,\sigma} \varepsilon_{i\alpha} \, \hat{n}_{i\alpha\sigma}
+ \sum_{i \ne j} \sum_{\alpha,\beta,\sigma} 
t^{\alpha\beta}_{ij} \, \hat{c}^\dagger_{i\alpha\sigma} \hat{c}_{j\beta\sigma} \nonumber \\
&\quad + \sum_{i,\alpha\beta} \sum_{\sigma\sigma^\prime}
U^{\sigma\sigma^\prime}_{\alpha\beta} \hat{n}_{i\alpha\sigma} \hat{n}_{i\beta\sigma^\prime},
\label{eq:Hamiltonian}
\end{align}
where $\hat{c}^\dagger_{i\alpha\sigma}$ ($\hat{c}_{i\alpha\sigma}$) creates (annihilates) an electron with spin $\sigma$ in orbital $\alpha$ at site $i$, and $\hat{n}_{i\alpha\sigma}=\hat{c}^\dagger_{i\alpha\sigma}\hat{c}_{i\alpha\sigma}$ is the number operator. The first two terms constitute the one-electron part of the Hamiltonian, comprising the on-site orbital energies and intersite hopping amplitudes. Fourier transformation of this one-electron Hamiltonian yields the non-interacting tight-binding Hamiltonian, $\hat{\mathcal{H}}^{\mathrm{tb}}_{0}(\mathbf{k})$, which serves as the DFT-derived input to the subsequent DMFT calculations.

The matrix element $U_{\alpha\beta}$ denotes the components of the static density-density interaction matrix spanning the active orbital indices ($\alpha, \beta$) and spin configurations ($\sigma, \sigma^\prime$) of the local multi-orbital subspace. Matching the electron-electron interaction terms defined in Eq.~\eqref{eq:Hamiltonian}, these elements take the explicit form:
\begin{equation}
    U^{\sigma\sigma^\prime}_{\alpha\beta} = 
    \begin{cases} 
      \frac{1}{2}U & \text{if } \alpha = \beta \text{ and } \sigma^\prime = \bar{\sigma}, \\
      \frac{1}{2}(U - 2J) & \text{if } \alpha \neq \beta \text{ and } \sigma^\prime = \bar{\sigma}, \\
      \frac{1}{2}(U - 3J) & \text{if } \alpha \neq \beta \text{ and } \sigma^\prime = \sigma,
    \end{cases}
\end{equation}
where $U$ signifies the intra-orbital double occupancy cost, $J$ denotes the Hund’s exchange coupling, and $\bar{\sigma}$ represents the opposite spin projection relative to $\sigma$.

The multi-orbital Hubbard model cannot, in general, be solved analytically in finite dimensions, while exact numerical approaches are limited by finite system sizes and computational cost. We therefore employ dynamical mean-field theory (DMFT), which maps the lattice problem onto a self-consistently embedded multi-orbital quantum impurity problem while treating itinerancy and local electronic correlations on an equal footing~\cite{Georges_RMP1996,Imada_RMP1998,Kotliar_RMP2006,Basov_RMP2011}. The retarded, local Green's function is given by
\begin{equation}
\hat{G}_{\text{loc}}(\omega^+) = \frac{1}{N_{\mathbf{k}}} \sum_{\mathbf{k}} 
\left[ (\omega^+ + \mu)\, \mathbb{I} - {\mathcal{\hat{H}}^{\mathrm{tb}}_{0}}(\mathbf{k}) - \hat{\Sigma}_{\text{imp}}(\omega^+) \right]^{-1},
\label{eq:gloc}
\end{equation}
where $\omega^+=\omega+i\eta$ with $\eta\rightarrow 0^+$. $\hat{\mathcal{H}}^{\mathrm{tb}}_{0}(\mathbf{k})$ denotes the Wannier-interpolated non-interacting Hamiltonian containing both intra-unit-cell hybridizations and intersite hopping amplitudes. Within the single-site DMFT approximation, the lattice self-energy is assumed to be purely local, $\hat{\Sigma}(\mathbf{k},\omega^+)=\hat{\Sigma}_{\mathrm{imp}}(\omega^+)$. The self-energy is expressed in the localized Wannier basis spanning the correlated subspace, allowing the lattice Green's function to be evaluated directly from the Wannier Hamiltonian and reducing the lattice problem to a self-consistently embedded quantum impurity coupled to an effective electronic bath.

\subsection{Multi-orbital iterative perturbation theory}

To solve the quantum impurity problem, we employ the multi-orbital iterated perturbation theory (MO-IPT)~\cite{Dasari_EPJB2016}. MO-IPT constructs the impurity self-energy from a second-order perturbative expansion around the Hartree limit and yields spectral and response functions directly on the real-frequency axis at all temperatures, including $T=0$. 
The MO-IPT self-energy ansatz is given by
\begin{equation}
\hat\Sigma_{\mathrm{imp}, \alpha}(\omega) = \sum_{\substack{\gamma \neq \alpha}} U_{\alpha\gamma} \, \langle \hat{n}_\gamma \rangle 
+ \frac{A_\alpha \sum\limits_{\substack{\gamma \neq \alpha}} \Sigma^{(2)}_{\alpha\gamma}(\omega)}
{1 - B_\alpha \sum\limits_{\substack{\gamma \neq \alpha}} \Sigma^{(2)}_{\alpha\gamma}(\omega)},
\label{eq:ansatz}
\end{equation}
This self-energy is diagonal in the orbital basis. The first term is the Hartree contribution, while the second incorporates the second-order pair-bubble self-energy,
\begin{multline}
    \Sigma_{\alpha\beta}^{(2)}(\omega) = U_{\alpha\beta}^2 \int\int\int d\epsilon_1 \, d\epsilon_2 \, d\epsilon_3 \, \rho_\alpha(\epsilon_1)\rho_\beta(\epsilon_2)\rho_\beta(\epsilon_3) \\
    \times \frac{n_{\mathrm{F}}(-\epsilon_1)n_{\mathrm{F}}(\epsilon_2)n_{\mathrm{F}}(-\epsilon_3) + n_{\mathrm{F}}(\epsilon_1)n_{\mathrm{F}}(-\epsilon_2)n_{\mathrm{F}}(\epsilon_3)}{\omega^+ - \epsilon_1 + \epsilon_2 - \epsilon_3},
\end{multline}
where $n_{\mathrm F}$ is the Fermi--Dirac distribution function, and $\rho_\alpha(\epsilon)=-(1/\pi)\,\mathrm{Im}\,\tilde{\mathcal G}_\alpha(\epsilon^+)$ is the local host spectral density. Here, $\tilde{\mathcal G}_\alpha$ denotes the Hartree-corrected bath propagator obtained from the Dyson-like relation
\begin{equation}
    \tilde{\mathcal{G}}_\alpha^{-1} = \left(\hat G_{\mathrm{loc}}^{-1} + \hat\Sigma + \hat\epsilon - (\mu - \mu_0)\mathbb{I} \right)_{\alpha\alpha}.
\end{equation}
The pseudo chemical potential, $\mu_{0}$ , is found at T = 0 by satisfying the Luttinger’s theorem or equivalently the Friedel’s sum rule, 
\begin{equation}
    -\frac{1}{\pi}{\rm Im}\int_{-\infty}^{+\infty} {\rm Tr}\Big(\frac{d\hat\Sigma(\omega)}{d\omega}\hat G_{loc}(\omega)\Big)d\omega = 0
\end{equation}
At ﬁnite temperature, an ambiguity exists in the determination of the pseudo-chemical potential. We choose to use the $\mu_{0}$ determined at zero temperature for all ﬁnite temperatures. The chemical potential, $\mu$, is found by ﬁxing the total occupancy from the local Green’s function,
$\hat G_{loc}$ , to be equal to the desired ﬁlling,
\begin{gather}
    -\frac{1}{\pi}{\rm Im}\int_{-\infty}^{+\infty} {\rm Tr}\,\hat G_{loc}(\omega)\,n_{f}(\omega)\,d\omega = n_{tot}
\end{gather}
where the trace is over the spin and orbital indices. The unknown coefficients $A_\alpha$, $B_\alpha$ from Eq.~\eqref{eq:ansatz} are obtained in the standard way by satisfying the high frequency limit and the atomic limit, respectively~\cite{Dasari_EPJB2016} as:
\begin{align}
    A_\alpha & = \frac{\sum_{\beta\ne\alpha} U^{2}_{\alpha\beta}\, \langle n_{\beta} \rangle(1-\langle n_\beta\rangle)}{\sum_{\beta\ne\alpha}U^{2}_{\alpha\beta} \, \langle n_{0\beta} \rangle\, (1-\,\langle n_{0\beta}\rangle)} \\
    + & \frac{\sum_{\beta\ne\alpha} U_{\alpha\beta}\sum_{\gamma\ne\beta\ne\alpha} U_{\alpha\gamma}(\langle n_{\beta}n_{\gamma}\rangle - \langle n_{\beta}\rangle \,\langle n_{\gamma} \rangle)}{\sum_{\beta\ne\alpha}U^{2}_{\alpha\beta} \, \langle n_{0\beta} \rangle\, (1-\,\langle n_{0\beta}\rangle)}.\\
B_{\alpha} 
&= \frac{
\mu_{0} + \epsilon_{\alpha} - \mu 
- \sum_{\beta \neq \alpha} U_{\alpha\beta} \langle n_{\beta} \rangle
}{\tau_{\alpha}} \nonumber \\ 
- & \frac{1}{\tau_{\alpha}^{2} A_{\alpha}}
\sum_{\{\beta,\gamma,\eta\} \neq \alpha}\!\!\!\!\!
U_{\alpha\beta} U_{\alpha\gamma} U_{\alpha\eta}
%&\qquad \times
\left[
\langle n_{\beta} \rangle \langle n_{\gamma} n_{\eta} \rangle
- \langle n_{\beta} n_{\gamma} n_{\eta} \rangle
\right].
\end{align}
The quantities $n_{\beta}$ and $n_{0\beta}$ denote the occupancies of orbital $\beta$ obtained from the full interacting Green's function and the Hartree-corrected bath Green's function, respectively. The parameter $\tau_{\alpha}$ is defined as
\begin{equation}
    \tau_{\alpha} = \sum_{\beta \ne \alpha} U_{\alpha\beta}^{2}\,
    \langle n_{0\beta} \rangle \left(1 - \langle n_{0\beta} \rangle \right).
\end{equation}
The quantities $\langle n_{\beta} \rangle$, $\langle n_{\beta} n_{\gamma} \rangle$, and $\langle n_{\beta} n_{\gamma} n_{\eta} \rangle$ represent one-, two-, and three-particle local density correlators evaluated within DMFT. To evaluate the two-particle correlation functions, we employ an approach based on the equation-of-motion formalism~\cite{Dasari_EPJB2016}, which yields the following exact constraint:
\begin{equation}
\sum_{m' \neq m} U_{mm'} \langle n_m n_{m'} \rangle = -\frac{1}{\pi} \int d\omega n_F(\omega) \mathrm{Im} \left[ \Sigma_m(\omega) G_m(\omega) \right].
\label{eq:two_particle_sum}
\end{equation}
Because this single relation couples multiple degrees of freedom, it cannot uniquely isolate each individual two-particle correlator on its own. To overcome this underdetermined system, we introduce a symmetric decoupling approximation that yields a closed-form expression for the individual orbital components:
\begin{equation}
\langle n_m n_{m'} \rangle = -\frac{\int d\omega n_F(\omega) \mathrm{Im} \left[ \Sigma_m(\omega) G_m(\omega) \right]}{\pi U_{mm'} (N_{\mathrm{orb}} - 1)},
\label{eq:two_particle_approx}
\end{equation}
for $U_{mm'}\neq 0$.
In the present implementation, three-particle correlators are neglected, while lower-order correlators are treated explicitly.

\subsection{Brillouin zone sums}
Accurate evaluation of Brillouin-zone integrals is essential for constructing the density of states and the local Green's function entering the DMFT self-consistency. To faithfully resolve sharp band-structure features such as van Hove singularities (vHs), we employ the momentum-dependent adaptive smearing scheme of Ref.~\cite{Yates_PRB2007}, which suppresses artificial broadening near stationary points of the electronic dispersion. As an illustration, the total density of states is defined as
\begin{equation}
D(\epsilon) = \sum_{\mathbf{k}}\sum_{\alpha} \delta(\epsilon - \epsilon_{\mathbf{k}\alpha}),
\end{equation}
where $\epsilon_{\mathbf{k}\alpha}$ are eigenvalues of $\hat{\mathcal{H}}^{\mathrm{tb}}_{0}(\mathbf{k})$. For numerical evaluation, the delta function is replaced by the product of a Gaussian and a second-order Hermite polynomial, namely, 
\begin{multline}
\delta(\epsilon - \epsilon_{\mathbf{k}}) \rightarrow 
\frac{1}{\sqrt{2\pi}w_{\vec k}}
\exp\left( -\frac{(\epsilon-\epsilon_{\vec k})^2}{2w^2_{\vec k}}\right) \\
\left(a_0+a_1H_2 \left (\frac{\epsilon-\epsilon_{\vec k}}{\sqrt2w_{\vec k}} \right) \right),
\end{multline}
where the adaptive broadening width is
\begin{equation}
w_{\mathbf{k}} = a \, |\nabla_{\mathbf{k}} \epsilon_{\mathbf{k}}|,
\label{eq:blochwidth}
\end{equation}
where $a\sim\mathcal O(1)$ is a dimensionless parameter.
This choice exploits the vanishing band velocity near vHS, ensuring minimal artificial broadening of sharp spectral features~\cite{Yates_PRB2007}. The band energies $\epsilon_{\mathbf{k}}$ are obtained by diagonalizing $\hat{\mathcal{H}}^{\mathrm{tb}}_{0}(\mathbf{k})$,  using the determinant condition $\left| \hat{\mathcal{H}}^{\mathrm{tb}}_{0}(\mathbf{k}) - \epsilon_{\mathbf{k}} I \right| = 0$. The gradients required for the adaptive broadening are obtained by differentiating the eigenvalue equation with respect to $k_m$ and using the determinant identity
\begin{equation}
\frac{d}{dx} \left| \hat{A}(x) \right| = \sum_{ij} \frac{d (A_{ij}(x))}{dx} \, C_{ij}(x),
\end{equation}
where $C_{ij}(x)$ are the elements of the cofactor matrix of $\hat A(x)$. Applying this identity yields an expression for $\nabla_{\mathbf k}\epsilon_{\mathbf k}$ in terms of the cofactor matrix of $[\hat{\mathcal{H}}^{\mathrm{tb}}_{0}(\mathbf{k})-\epsilon_{\mathbf k}I]$.
The momentum derivatives of the Hamiltonian are evaluated directly using its Fourier representation,
\begin{equation}
\frac{\partial \hat{\mathcal{H}}^{\mathrm{tb}}_{0}(\mathbf{k})}{\partial k_m}
= \sum_{\mathbf{R}} i R_m \, e^{i \mathbf{k}\cdot\mathbf{R}} \hat{\mathcal{H}}^{\mathrm{tb}}_{0}(\mathbf{R})), \, (m=x,y,z)
\end{equation}
which allows computation of $\nabla_{\mathbf{k}} \epsilon_{\mathbf{k}}$ without finite-difference approximations and enables a stable implementation of the adaptive smearing scheme.

Brillouin-zone (BZ) summations required for evaluating the local density of states (DOS), Green's function, and transport properties are performed over the irreducible Brillouin zone (IBZ), with symmetry weights assigned to each $\mathbf{k}$-point to exploit the crystal symmetries and thereby reduce the computational cost~\cite{Monkhorst_PRB1976,Martin2004, Ong_2013}. This approach is particularly effective for cubic SrVO$_3$, where the crystal symmetry renders the three $t_{2g}$ bands nearly degenerate and symmetry equivalent. Consequently, the number of $\mathbf{k}$-points can be substantially reduced while maintaining numerical accuracy. In practice, we have used a $30\times30\times30$ mesh, corresponding to 816 irreducible $\mathbf{k}$-points for a smearing parameter of $a=1.0$. The complete $t_{2g}$ manifold is then reconstructed by combining three symmetry-equivalent IBZs, yielding a total of $3\times816=2448$ $\mathbf{k}$-points. The uniform broadening parameter, $\eta$ that appears as $\omega+i\eta$ e.g.\ in Eq.~\eqref{eq:gloc} is replaced by a $\mathbf{k}$-dependent broadening, $\eta_{\mathbf{k}}$, which is just the adaptive smearing computed above (Eq.~\ref{eq:blochwidth}) in the Bloch basis, and rotated into the Wannier basis. All DMFT calculations are iterated to self-consistency until the change in the local Green's-function matrix between successive iterations is less than $10^{-4}$. The converged self-energies are subsequently used to compute the spectral and transport properties discussed in Sec.~\ref{sec:results}. Next, we describe the transport formalism employed in this work.

\subsection{Transport}
The optical conductivity is evaluated within linear-response theory using the Kubo formalism~\cite{Tomczak_PhD,Basov_RMP2011}. Within DMFT, vertex corrections vanish identically for the single-band Hubbard model in the infinite-dimensional limit, allowing the optical conductivity to be evaluated exactly from the current-current bubble constructed from the fully interacting Green's function~\cite{Pruschke_PRB1993,Georges_RMP1996}. For multi-orbital systems, however, this exact cancellation is generally not guaranteed because of inter-orbital contributions to the current operator. In the present work, following the widely adopted approximation employed in many DFT+DMFT studies of correlated materials~\cite{Kotliar_RMP2006,Tomczak_PhD,Tomczak2009,Si2021}, we evaluate the optical conductivity within the current-current bubble approximation using the fully interacting Green's function. Although vertex corrections can quantitatively modify the optical conductivity in finite-dimensional and multi-orbital systems~\cite{Sadovskii2007,Sato_PRB2016, Rohringer_RMP2018,Mu_PRB2022}, their evaluation lies beyond the scope of the present MO-IPT implementation.

Real-frequency impurity solvers such as MO-IPT enable the direct evaluation of transport properties without analytic continuation. Within the local approximation, the real part of the optical conductivity along the x-direction is given by \cite{Georges_RMP1996,Pruschke_PRB1993,Held2007,Tomczak2009}:
\begin{multline}
    \sigma_{1}(\Omega) = \sigma_0  \int_{\mathrm{BZ}} \frac{d^3k}{8\pi^2} \int_{-\infty}^{\infty} d\omega\, \frac{n_f(\omega) - n_f(\omega+\Omega)}{\Omega} \\ 
    \mathrm{Tr}\!\left[ \hat{\nu}^{x}(\mathbf{k}) \hat{A}(\mathbf{k}, \omega+\Omega) \hat{\nu}^{x}(\mathbf{k}) \hat{A}(\mathbf{k}, \omega) \right],
    \label{eq:sigma_trace}
\end{multline}
where $\hat{\nu}^{x}(\mathbf{k})$  represents the velocity matrix operator, $\hat{A}(\mathbf{k}, \omega)$ is the matrix-valued interacting spectral function, and the trace ($\mathrm{Tr}$) runs explicitly over the active orbital indices of the correlated subspace, $n_f(\omega)$ is the Fermi-Dirac distribution function, and $\sigma_0$ is a material-dependent constant that is obtained in this work through a comparison of theory with the experimental DC resistivity.

The DC conductivity follows from the $\Omega \to 0$ limit of Eq.~\eqref{eq:sigma_trace}:
\begin{equation}
\sigma_{\mathrm{DC}} = 
\sigma_0
\int\limits_{\substack{\mathrm{BZ}}}\frac{d^3k}{8\pi^2}
\int d\omega
\left(-\frac{\partial n_f}{\partial \omega}\right)
\!\mathrm{Tr}\left[
\Big(\hat{\nu}^{x}(\mathbf{k}) \hat{A}(\mathbf{k},\omega)\Big)^{2}
\right].
\label{eq:sigma_DC}
\end{equation}
The formalism outlined in this section is employed to compute the spectral and transport properties of SrVO$_3$. The resulting calculations and their comparison with experiment are presented and discussed in the following section.

\section{Results and Discussion}
\label{sec:results}

SrVO$_3$ is a prototypical moderately correlated metal with a nominal V$^{4+}$ ($3d^1$) electronic configuration, in which the single electron occupies the triply degenerate $t_{2g}$ manifold~\cite{Fujimori_PRL1992,Fujimori_PRB1995}. Owing to the ideal cubic symmetry, the three $t_{2g}$ orbitals remain symmetry equivalent, resulting in negligible orbital polarization and nearly identical orbital occupations~\cite{Nekrasov_PRB2005,Taranto_PRB2013}. Consequently, the DMFT self-energy is orbital diagonal and nearly identical across the three $t_{2g}$ orbitals.
\begin{equation*}
\hat{\Sigma}_{\alpha\beta}(\omega) = \Sigma(\omega)\delta_{\alpha\beta}.
\end{equation*}
Early photoemission measurements revealed a coherent quasiparticle peak at the Fermi level accompanied by incoherent Hubbard-band features, indicating substantial but not overwhelming correlation effects~\cite{Fujimori_PRL1992,Fujimori_PRB1995}. Subsequent ARPES studies established a quasiparticle mass enhancement of approximately $m^*/m \simeq 2$, corresponding to a quasiparticle weight $Z \simeq 0.5$~\cite{Sekiyama_PRL2004,Yoshida_PRL2005,Aizaki_PRL2012}. Material-specific DFT+DMFT calculations likewise reproduced the experimentally observed bandwidth renormalization and consistently reported quasiparticle weights in the range $Z \approx 0.4$--$0.6$~\cite{Nekrasov_PRB2005,Yoshida_PRL2005,Aizaki_PRL2012,Taranto_PRB2013}.
Taken together, these experimental and theoretical results indicate that the low-energy electronic structure of SrVO$_3$ is governed by a single quasiparticle coherence scale, making it a suitable platform for investigating quasiparticle renormalization and adiabatic continuity in a realistic multi-orbital setting.

\subsection{Adiabatic continuity and quasiparticle renormalization}
\label{sec:adiabatic}
Motivated by the experimentally established Fermi-liquid behavior of SrVO$_3$, we expand the local self-energy to linear order in frequency~\cite{Leggett_RMP1975,Abrikosov_Book1975,GDMahan_Book2000},
\begin{equation}
    \Sigma(\omega^+) 
    \simeq \Sigma(0)
    + (1 - Z^{-1})\,\omega + \mathcal{O}(\omega^{2}, T^{2}) \,,
    \label{eq:sigfl}
\end{equation}
For the orbital-diagonal self-energy discussed above, this gives
\begin{equation}
    \hat{\Sigma}(\omega) \simeq \big[ \Sigma(0)
    + (1 - Z^{-1})\,\omega\big] I \, ,
    \label{eq:selfexp}
\end{equation} where the quasiparticle weight is defined as
    \begin{equation}
    Z 
    = \left[1 - 
    \left.\frac{\partial {\rm Re}\Sigma(\omega)}{\partial\omega}\right|_{\omega=0}
    \right]^{-1} ,
    \label{eq:QP}
\end{equation}
Substituting Eq.~\eqref{eq:selfexp} into Eq.~\eqref{eq:gloc} yields the quasiparticle Green's function,
\begin{equation}
    G_{\mathrm{latt}}^{\mathrm{qp}}(k,\omega^+)
    \simeq 
    \left[\left(\frac{\omega^+}{Z} + \mu - \Sigma(0)\right)I + i\hat\eta(k) - \hat{H}_0(k)\right]^{-1}.
    \label{eq:Gqp}
\end{equation}
Eq.~\eqref{eq:Gqp} shows that electronic correlations renormalize the frequency by the quasiparticle weight $Z$ and shift the chemical potential by $\Sigma(0)$, while preserving the functional form of the non-interacting Green's function~\cite{Georges_RMP1996,Toschi_PRB2005}. The corresponding quasiparticle dispersion becomes
\begin{equation}
    \tilde{\varepsilon}_{k} = Z\big[\varepsilon_{k} - \mu + \Sigma(0)\big],
\end{equation}
leading to the familiar mass enhancement relation $m^{\ast}/m = Z^{-1}$. Thus, within the Fermi-liquid regime, the low-energy electronic structure is governed by a single scalar quasiparticle renormalization factor, $Z$, implying that the interacting spectrum is obtained from its non-interacting counterpart through a uniform rescaling of the energy axis. In more general multiorbital systems without cubic symmetry or orbital degeneracy, no single quasiparticle renormalization factor can map the interacting Green's function onto its non-interacting counterpart. Instead, different orbitals may renormalize with distinct coherence scales, giving rise to band-dependent mass enhancements~\cite{Tamai_PRX2019}, orbital-selective quasiparticle dynamics~\cite{Yi_PRL2013}, and a breakdown of the simple adiabatic mapping described above~\cite{Medici_PRL2011}.  

This observation forms the central hypothesis of the present work: if the low-energy physics of SrVO$_3$ is governed predominantly by the quasiparticle weight, then different interaction parameters yielding the same $Z$ should produce nearly identical low-energy spectral and transport properties. The following sections test this hypothesis quantitatively.

\subsection{$U-J$ parameter space expolaration}
\label{sec:contour}

\begin{figure}[htbp]
    \centering
    \includegraphics[width=0.48\textwidth]{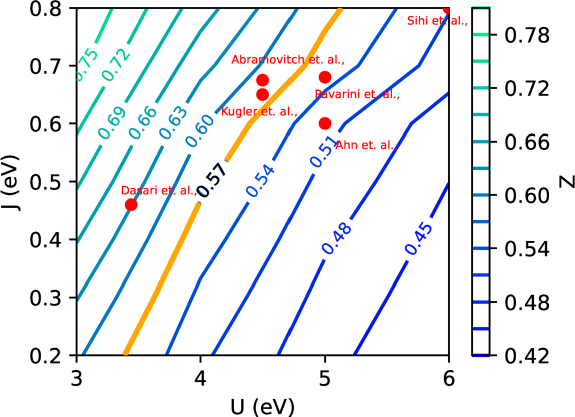}
    \caption{Contour map of the quasiparticle weight $Z$ in the $U$--$J$ plane for SrVO$_3$, computed within DFT+DMFT using the MO-IPT impurity solver. The solid curves denote iso-$Z$ contours, with the corresponding quasiparticle weights indicated by the legend and the color bar.}
    \label{fig:Z_contour}
\end{figure}

Although the single-particle Hamiltonian is fixed by the DFT-based Wannier construction, the interaction parameters of the multi-orbital Hubbard Hamiltonian are not known \textit{a priori} and have been the subject of extensive theoretical investigation for SrVO$_3$. Early constrained density-functional and constrained random-phase approximation (cRPA) calculations reported Hubbard interactions in the range $U\sim3-5$~eV and Hund's couplings $J\sim0.6-0.8$~eV, depending on the choice of correlated subspace and screening channels~\cite{Aryasetiawan_PRB2006,Miyake_PRB2008}. Subsequent DFT+DMFT studies have employed similar interaction strengths. We have marked some of the parameter sets used in the literature~\cite{Pavarini_PRL2004, Dasari_EPJB2016, GAhn_PRB2022, Abramovitch2024, Kugler_PRB2026, sihi2026} on the $U-J$ plane in Fig.~\ref{fig:Z_contour}. These studies employ a diverse range of interaction parameters, yet successfully reproduce key aspects of the electronic structure, including quasiparticle renormalization, bandwidth reduction, and Hubbard-band formation~\cite{Nekrasov_PRB2005,Yoshida_PRL2005,Tomczak_EPL2012,Taranto_PRB2013}. This leads us to speculate that a quantitative description of the properties of SrVO$_3$ is weakly sensitive to the choice of interaction parameters as long as the parameters yield the same quasiparticle weight (or equivalently, mass enhancement).
In order to examine this hypothesis, we carry out a comprehensive exploration of the $(U,J)$ parameter space within the DFT+DMFT framework at $T=0$, and mark the iso-$Z$ contours where $Z$ is the quasiparticle weight. For a representative iso-$Z$ contour, we then examine the sensitivity of the transport properties to the choice of interaction parameters. Such a systematic exploration is computationally feasible because, unlike CTQMC, the computational cost of the MO-IPT impurity solver scales polynomially with the number of orbitals and depends only weakly on increasing interaction strength or decreasing temperature. The on-site Hubbard repulsion and Hund's coupling are varied over the intervals $U = 3.0-6.0~\mathrm{eV}$ and $J = 0.2-0.8~\mathrm{eV}$, respectively. For each $(U,J)$ pair, the real-frequency impurity self-energy obtained from MO-IPT is used to compute the quasiparticle weight $Z(U,J)$ and the corresponding lattice Green's function.

The resulting $Z(U,J)$ contour map, illustrated in Fig.~\ref{fig:Z_contour}, reveals the competing influence of $U$ and $J$. Each solid line represents an iso-$Z$ curve, and the legend represents the value of the quasiparticle weight specific to that curve. As expected, increasing $U$ for a fixed $J$ systematically suppresses quasiparticle weight. Increasing $J$ at fixed $U$ reduces the effective inter-orbital interaction strengths, which enter as $U-2J$ and $U-3J$ for opposite- and parallel-spin configurations, respectively. This leads to a decrease in correlation strength and an increase in quasiparticle weight. Consequently, the iso-$Z$ contours exhibit a positive slope in the $(U,J)$ plane, reflecting the compensating effects of increasing $U$ and $J$ on the overall correlation strength.

\subsection{Fermi-liquid self-energies}
\label{sec:FLSE}

Having established the quasiparticle renormalization landscape, we now use the iso-$Z$ contours as an organizing framework for the subsequent analysis. Among the interaction parameters employed in the recent literature, Kugler \textit{et al.}~\cite{Kugler_PRB2026} used $U=4.5$ eV and $J=0.65$ eV, for which the present MO-IPT calculations yield a quasiparticle weight of $Z\approx0.57$. We therefore focus on the corresponding iso-$Z$ contour (orange curve in Fig.~\ref{fig:Z_contour}) to investigate how the self-energy and transport properties evolve with different $(U,J)$ combinations while maintaining a fixed quasiparticle renormalization.

\begin{figure}[htbp]
    \centering
    \includegraphics[scale=0.5]{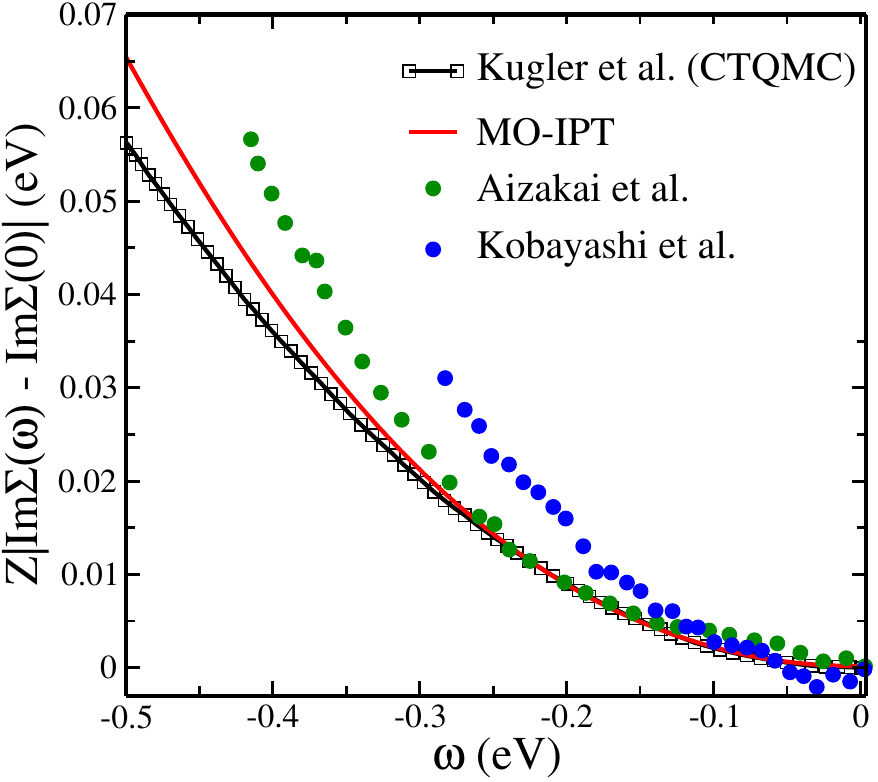}
    \caption{Comparison of the real-frequency self-energy of SrVO$_3$ obtained from DFT+DMFT using the MO-IPT impurity solver (red line) with the CTQMC results of Ref.~\cite{Kugler_PRB2026} (black squares) for $U=4.5$~eV, $J=0.65$~eV, and $T=116$~K. Also shown are the self-energies from the experimental ARPES momentum distribution curve from Refs.~\cite{Aizaki_PRL2012,Kobayashi2017} (green and blue symbols), measured below $20$~K.}
    \label{fig:FLSE}
\end{figure}

Fig.~\ref{fig:FLSE} compares the calculated self-energy against experiment and numerically exact CTQMC calculations, providing a benchmark of the MO-IPT approach. The quantity $Z|\mathrm{Im}\Sigma(\omega)-\mathrm{Im}\Sigma(0)|$, calculated within DFT+DMFT using the MO-IPT solver for $U=4.5$ eV, $J=0.65$ eV and $T=116$ K, is compared with the corresponding experimental self-energy extracted from ARPES momentum-distribution curves (MDCs)~\cite{Aizaki_PRL2012,Kobayashi2017}. We also compare with CTQMC results reported by Kugler \textit{et al.}~\cite{Kugler_PRB2026}, obtained for the same interaction parameters at $T=116$ K, with the reported quasiparticle weight $Z\approx0.50$.

The MO-IPT self-energy (solid red line) exhibits the characteristic Fermi-liquid behavior, $\mathrm{Im}\Sigma(\omega)\propto\omega^2$ at low frequencies, with a curvature consistent with the quasiparticle renormalization factor $Z\approx0.57$ extracted from the low-frequency slope of $\mathrm{Re}\Sigma(\omega)$~\cite{Georges_RMP1996}. Our result is in good agreement with the experimental self-energy (solid circles) extracted from momentum distribution curve (MDC) analysis of ARPES data, as well as the CTQMC data (squares, ~\cite{Kugler_PRB2026}) indicating that the quasiparticle scattering rate follows the expected $\omega^2$ scaling in the Fermi-liquid regime of SrVO$_3$. The comparison also shows that the real-frequency MO-IPT framework provides a reliable description of the low-energy electronic dynamics.

We further find that the self-energy exhibits an approximately universal frequency dependence over the low-energy range considered. As demonstrated in Fig.~\ref{fig:FLSE_scaling} in Appendix~\ref{sec:appendix_A}, the quantity $Z|\mathrm{Im}\Sigma(\omega)|$ computed within MO-IPT at $T=0$ for several $(U,J)$ combinations lying on the iso-$Z$ contour corresponding to $Z=0.57$ nearly collapses onto a single curve. This behavior indicates that the low-energy frequency dependence of the self-energy is governed primarily by the quasiparticle renormalization factor, with only a weak dependence on the individual interaction parameters. The overall agreement with both experiment and CTQMC demonstrates that MO-IPT captures the essential low-energy many-body dynamics of SrVO$_3$. This approximate universal scaling naturally motivates the transport analysis that follows, where we examine whether the same quasiparticle coherence scale likewise governs the dc and optical transport properties.

\subsection{DC resistivity}
\label{sec:dc}

Recent experimental and theoretical studies have demonstrated $\mathrm{SrVO}_3$ as a clean benchmark system for correlated-metal transport, where electron-electron ($e$-$e$) and electron-phonon ($e$-$ph$) scattering contributions may be quantitatively disentangled~\cite{Brahlek2024,Abramovitch2024,Kugler_PRB2026}. In particular, ultra-clean stoichiometric samples exhibit Fermi-liquid transport at low temperatures, while $e$-$ph$ scattering was shown to be dominant over a wide intermediate-to-high temperature range~\cite{Brahlek2024,Abramovitch2024,Kugler_PRB2026}. 
 
\begin{figure}[htbp] 
\centering
\includegraphics[scale=0.5]{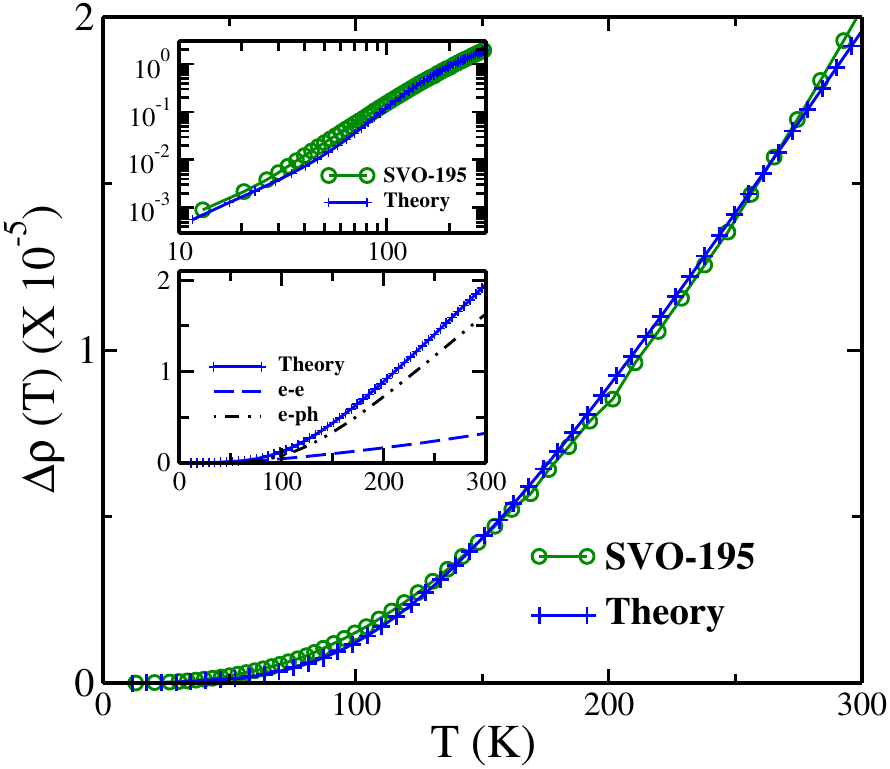}
\caption{Comparison of the measured dc resistivity of ultraclean SrVO$_3$ (green circles) with the calculated resistivity (plus symbols), showing good agreement over the entire temperature range. The upper inset replots the same data on a log--log scale to emphasize the low-temperature behavior. The lower inset decomposes the calculated resistivity into electron--electron ($e$--$e$, dashed line) and electron--phonon ($e$--ph, dash--dot--dot line) contributions, together with the total resistivity (solid line). The two scattering mechanisms contribute equally near $\sim80$~K, marking the crossover from electron--electron- to electron--phonon-dominated transport.}
\label{fig:rho}
\end{figure}

Since the multi-orbital Hubbard Hamiltonian (Eq.~\eqref{eq:Hamiltonian}) includes only $e-e$ interactions, we use a phenomenological approach, namely the Bloch-Gr\"uneisen (BG) expression~\cite{Inoue_PRB1998, Cvijovic2011, AshcroftMermin, Ziman1960} to incorporate $e-ph$ interactions. The contribution to resistivity due to $e-ph$ scattering within the BG approach is given by:
\begin{equation}
\rho_{e-ph}(T) = \rho_R \left(\frac{T}{\Theta_R}\right)^n \int_{0}^{\Theta_R/T} \frac{x^n}{(e^x - 1)(1 - e^{-x})} \, dx,
\end{equation}
where $\rho_R$ is a constant prefactor that depends on the strength of electron-phonon coupling, and $\Theta_R$ is the characteristic Bloch-Gr\"uneisen (transport) temperature. We observe that the the above expression with $n=5$ and $\theta_R\sim 700$ fits the first-principles DFPT $e-ph$ calculations from Ref.~\cite{Abramovitch2024} quite well. These fitting parameters are strikingly close to the   
ones found phenomenologically in a previous work~\cite{Inoue_PRB1998}. The $e-e$ contribution to the temperature-dependent dc resistivity is computed from the zero-frequency limit of the optical conductivity: $\rho_{e-e}(T) = 1/\sigma_{\mathrm{dc}}(T)$ using Eq.~\eqref{eq:sigma_DC}. So, the theoretically computed dc resistivity is a sum of the $e-e$ and $e-ph$ contributions:
\begin{equation}
    \rho_{\rm Theory}(T)=\rho_{e-e}(T) + \rho_{e-ph}(T)
    \label{eq:totrho}
\end{equation}
The unknown parameters in the above equation are $n, \theta_R, \rho_R$ and $\sigma_0$. 
In this work, we fix $n=5$ and use the prefactors $\rho_R$, and $\sigma_0$ (Eq.~\ref{eq:sigma_DC}) as fitting parameters for comparing theory (Eq.~\eqref{eq:totrho}) with experimental data for ultraclean SrVO$_3$~\cite{Brahlek2024}. We find that the temperature scale $\theta_R=700$K yields a negative  $\sigma_0$, which is unphysical. Hence, we have varied $\theta_R$, and found that the best fit between theory and experiment was obtained for $\theta_R=800$K. The prefactors were found to be $\rho_R=2.5\times 10^{-4}\, \Omega\,{\rm cm}$
and $\sigma_0=1.21\,\Omega^{-1}\,{\rm cm}^{-1}$, where the latter has been used in computing the optical conductivity as well. 

The main panel of Fig.~\ref{fig:rho} displays the calculated $\rho_{\rm Theory}(T)$ (labeled as `Theory') and the resistivity data for ultraclean SrVO$_3$ digitized from Ref.~\cite{Brahlek2024} (labeled as `SVO-195'). The agreement between theory and experiment is seen to be reasonable over the entire temperature range. The upper inset shows the same data as the main panel albeit on a log-log scale to demonstrate that the good agreement extends to the lowest temperature scales. The Fermi liquid $T^2$ part present in the experimental data below about $25$ K is captured very well by the theory, and as expected, arises solely through the $e-e$ contribution. The lower inset shows the $e-e$ and $e-ph$ contributions to the total dc resistivity. Although it is not visible from the plot, we find that the two contributions cross each other around $80$ K, with the former being dominant for lower temperatures, and the latter for higher temperatures. Although the calculations have been performed for a fixed set of parameters, namely $U=4.5$ eV and $J=0.65$ eV, we show in Appendix ~\ref{sec:appendix_A} that the dc resistivity in the range of $0-300$ K is hardly dependent on the specific $(U,J)$ combinations if they lie on a single iso-$Z$ contour (see Fig.~\ref{fig:rho_diff_z} and the associated text). 

The weak dependence of the dc resistivity on the individual interaction parameters further supports the picture that the low-temperature transport is governed primarily by the quasiparticle coherence scale. It is therefore natural to ask whether this approximate universality also extends to the frequency-dependent optical response. We now turn to the optical conductivity to examine this question and to benchmark the calculated spectra directly against experiment.

\subsection{Optical conductivity}
\label{sec:optics}
The optical conductivity is calculated using Eq.~\eqref{eq:sigma_trace} in the Wannier basis. The corresponding velocity matrix elements are evaluated within the Peierls approximation,
\begin{equation}
\hat{v}_{x}(\mathbf{k}) = \frac{1}{\hbar}
\frac{\partial \hat{\mathcal{H}}_0(\mathbf{k})}{\partial k_x},
\end{equation}
which retains only the Hamiltonian-derivative contribution to the velocity operator and neglects the additional gauge-covariant Berry-connection terms arising from the $\mathbf{k}$-dependence of the Wannier basis~\cite{Wissgott2012,Lorenzo2025}. This approximation is standard in DFT+DMFT optical conductivity calculations for transition-metal oxides with localized $d$ orbitals~\cite{Wissgott2012}.

\begin{figure}[htbp] 
    \centering
    \includegraphics[scale=0.5]{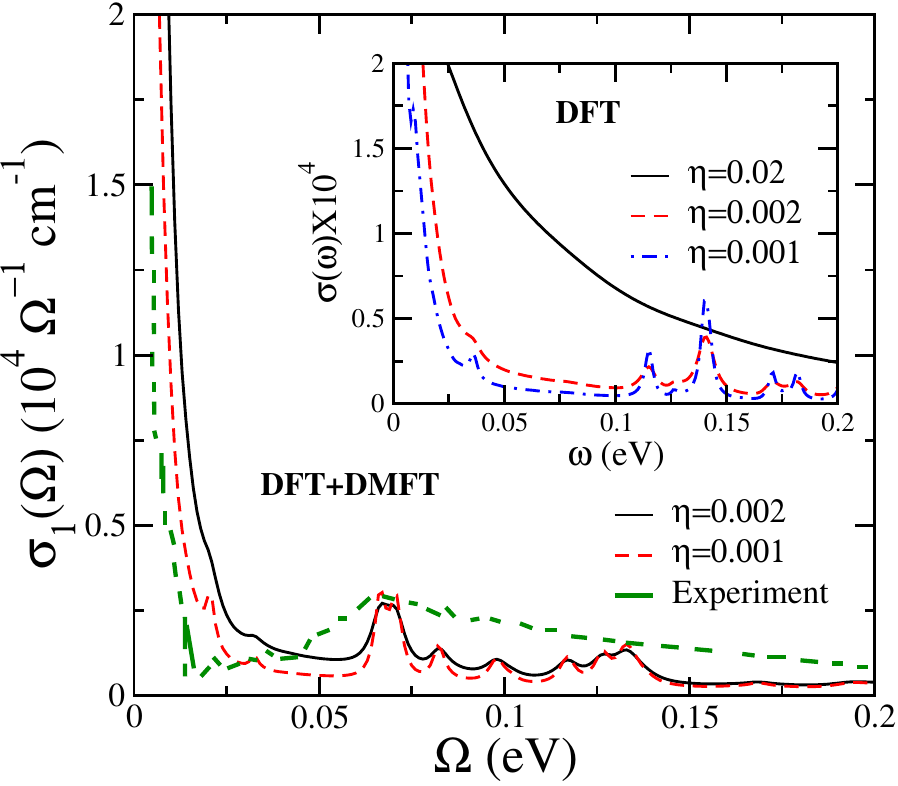}
    \caption{Optical conductivity of SrVO$_3$ calculated within DFT+DMFT using the MO-IPT impurity solver for $U=4.5$~eV and $J=0.65$~eV at $T\approx6$~K. Results for broadening parameters $\eta=0.002$ (black solid line) and $\eta=0.001$ (red dashed line) are compared with the experimental spectrum (green dashed line) of a highly stoichiometric sample with residual resistivity ratio $\mathrm{RRR}=130$, digitized from Ref.~\cite{GAhn_PRB2022}. Inset: Non-interacting optical conductivity calculated from the DFT Wannier Hamiltonian for several values of the broadening parameter $\eta$.}
    \label{fig:optics}
\end{figure}

Fig.~\ref{fig:optics} displays the optical conductivity computed using Eq.~\eqref{eq:sigma_trace} within the DFT+DMFT (MO-IPT) framework for the interaction parameters $U=4.5~\mathrm{eV}$ and $J=0.65~\mathrm{eV}$~\cite{Kugler_PRB2026}. The fully interacting theoretical optical conductivities obtained using broadening parameters $\eta=0.002$ (black line) and $\eta=0.001$ (red dashed line) are compared directly with the high-resolution experimental data digitized from Ref.~\cite{GAhn_PRB2022} for a very clean sample with residual resistivity ratio $\mathrm{RRR}=130$ measured at $T\approx6~\mathrm{K}$. 

The optical conductivity profile is characterized by two distinct low-energy features. In the limit $\Omega\ll W$, the spectrum is dominated by a coherent Drude response arising from intraband transitions within the correlated V $t_{2g}$ manifold. The calculated conductivity  agrees reasonably well with the magnitude and the characteristic line shape of the experimental Drude peak, indicating that the low-energy quasiparticle dynamics and scattering rates are captured within the chosen interaction regime.

At slightly higher frequencies, a pronounced interband feature emerges near $\Omega \approx 70~\mathrm{meV}$. Within our framework, this peak is identified as a direct consequence of the correlation-induced renormalization of interband excitations inherent to the underlying SrVO$_3$ band structure. Evidence for its one-particle origin is already visible in the noninteracting optical conductivity obtained from the DFT Wannier Hamiltonian, shown in the inset of Fig.~\ref{fig:optics}, where the corresponding interband feature appears near $0.14~\mathrm{eV}$. The inset also illustrates the effect of the numerical broadening parameter $\eta$, which controls the width of the broadened Dirac delta function used in evaluating the noninteracting optical conductivity. At $\eta = 0.02$~eV, the interband feature is completely washed out and only a broad Drude-like peak survives; as $\eta$ is reduced to $0.002$ and $0.001$~eV, the interband peak progressively emerges, indicating that $\eta$ effectively mimics the role of disorder-induced scattering in the system. As $\eta$ is reduced, the interband feature becomes progressively sharper while its peak position remains essentially unchanged, confirming that the feature is intrinsic to the underlying DFT Wannier band structure rather than a consequence of numerical broadening. Electronic correlations subsequently renormalize this energy scale to approximately $70~\mathrm{meV}$, producing the experimentally observed low-energy peak. As demonstrated in Appendix~\ref{sec:appendix_B} through a systematic analysis of the joint density of states, velocity matrix elements, and Fermi occupation factors, the associated spectral weight originates from the multiband structure of the $t_{2g}$ manifold. Importantly, the low-energy interband feature disappears when the off-diagonal hopping elements of the Wannier Hamiltonian are suppressed (see Fig.~\ref{fig:interband}), demonstrating that these hybridization channels are essential for its formation. The role of the off-diagonal hoppings is to generate the band splittings and nearly parallel dispersions that enhance the low-energy joint density of states and give rise to the observed optical transition. This finding is consistent with earlier studies that emphasized the importance of orbital off-diagonal hopping processes in the low-energy optical response of SrVO$_3$~\cite{GAhn_PRB2022}.

The computed interband peak, while correctly positioned near $\Omega \approx 70~\mathrm{meV}$, is noticeably narrower than its experimentally observed counterpart. This discrepancy could arise from multiple reasons. First, single-site DMFT neglects momentum-dependent (non-local) self-energy corrections, which could redistribute spectral weight and broaden interband features beyond what a purely local self-energy captures~\cite{GW1998,Maier_RMP2005,Sadovskii_PRB2005,Toschi2007,Held2007,Rohringer_RMP2018}. Second, residual disorder, which is present even in ultraclean SrVO$_3$ films as evidenced by a finite residual resistivity $\rho_0$, introduces an elastic scattering rate that uniformly broadens spectral features and is not included in our bulk periodic calculation. 
Finally, and most directly relevant given the energy scale involved, the electron-phonon ($e$-ph) interaction provides a natural broadening channel.
The associated $e$-ph self-energy $\Sigma_\text{ph}(\omega, T)$, entering through the Migdal-Eliashberg expression~\cite{Allen1971, Marsiglio2020}, produces phonon-assisted broadening of optical transitions via the Holstein mechanism~\cite{Allen1971}, with clear precedent in the mid-infrared optical conductivity of the structurally analogous perovskite SrTiO$_3$~\cite{vanMechelen2008, Devreese2010} and across correlated transition metal oxides more broadly~\cite{Basov_RMP2011, Okazaki2006}. In SrVO$_3$, $e$-ph coupling has been shown to contribute significantly to resistivity over a wide temperature range down to $\sim$30~K, and to induce observable kinks in the quasiparticle spectral function~\cite{Abramovitch2024}. Correlation effects further enhance the coupling to Jahn-Teller optical phonon modes of the VO$_6$ octahedra whose energy scale coincides directly with that of the interband feature~\cite{Abramovitch2025, Mirjolet2021}. Hence, we speculate that, for SrVO$_3$, $e$-ph coupling may be the dominant reason for the discrepancy. 

Having established that the MO-IPT framework quantitatively reproduces both the experimental optical line shape and the Fermi-liquid self-energy at the lowest measured temperature, we now turn to the temperature evolution of the optical conductivity. While the comparison in Fig.~\ref{fig:optics} represents a benchmark at a single temperature, namely, $T \approx 6~\mathrm{K}$, the temperature dependence of the optical response provides a more demanding test of the theory, probing how the interplay between quasiparticle coherence and electronic scattering evolves across the Fermi-liquid to incoherent-metal crossover. In particular, the fate of the low-energy interband feature, whose position and spectral weight are well captured at the base temperature and the evolution of the Drude response with increasing thermal fluctuations offer direct insight into the temperature scales governing quasiparticle decoherence in SrVO$_3$. So, we now turn to the full temperature and frequency dependence of the optical conductivity.

\begin{figure}[htbp] 
    \centering
    \includegraphics[scale=0.5]{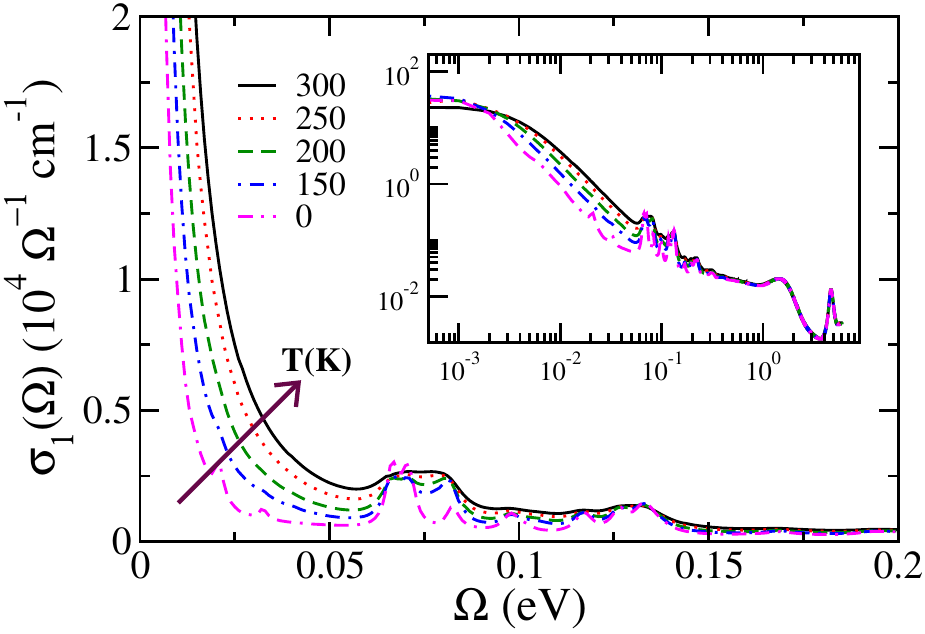}
    \caption{The main panel shows temperature dependence of the optical conductivity of SrVO$_3$ for $U=4.5$~eV and $J=0.65$~eV. With increasing temperature, the low-energy interband feature and the intraband (Drude) response progressively broaden and merge, making their separation increasingly indistinct. The inset replots the optical conductivity on a log--log scale over an extended energy range up to 6~eV, highlighting the mid-infrared feature near $\sim1.5$~eV and the higher-energy feature around $\sim5$~eV.}
    \label{fig:optics_T}
\end{figure}

Fig.~\ref{fig:optics_T} shows the temperature evolution of the theoretical optical conductivity of SrVO$_3$ computed within the DFT+DMFT framework. At low temperatures, as discussed previously in Fig.~\ref{fig:optics}, the optical spectrum exhibits a well-defined Drude response accompanied by a distinct low-energy interband transition feature, reflecting the presence of coherent quasiparticles in the correlated metallic state. As temperature increases, the low-energy interband peak progressively broadens and loses intensity, while the Drude response becomes increasingly diffuse. Consequently, the separation between intraband and interband contributions becomes less pronounced at elevated temperatures. This behavior is consistent with the gradual reduction of quasiparticle coherence due to enhanced electronic scattering, which transfers spectral weight over a broader energy range and smears out the low-energy optical features, very similar to what is seen in experiments~\cite{GAhn_PRB2022}. The $\Omega\rightarrow 0$ limit indicates that the system remains metallic throughout the temperature range considered, while undergoing a continuous crossover from a highly coherent low-temperature Fermi-liquid regime to a less coherent correlated metal at higher temperatures.

The inset of Fig.~\ref{fig:optics_T} shows the optical conductivity over a much larger frequency scale, where the conventional mid-infrared peak is observed at about 1.5 eV.
We note that it is important to distinguish the low-energy interband feature observed near $70~\mathrm{meV}$ in the optical conductivity of ultraclean SrVO$_3$ from the mid-infrared (MIR) peak conventionally discussed in the correlated-oxide literature. In single or few-orbital Hubbard-model calculations, as well as in realistic DFT+DMFT studies of correlated transition-metal oxides, the canonical MIR feature typically appears at an energy scale ${\cal{O}}(\rm eV)$, arising from spectral-weight transfer between the renormalized quasiparticle bands and the incoherent Hubbard bands~\cite{Imada_RMP1998,Basov_RMP2011}. This Hubbard-physics MIR scale, seen in the inset of Fig.~\ref{fig:optics_T}, at $\sim$1.5 eV, has a distinct origin from the $70~\mathrm{meV}$ feature in SrVO$_3$. The latter is determined not by the correlation strength but by the off-diagonal-hopping-induced splitting within the $t_{2g}$ manifold, which represents a one-particle band-structure effect that produces nearly parallel split bands along specific $\mathbf{k}$-space directions, generating a sharp peak in $\sigma(\omega)$ through a van-Hove-like accumulation of spectral weight in the joint density of states~\cite{GAhn_PRB2022}. We conclude the section on optics with a discussion of the optical lineshape on the interaction parameters. Similar to the finding for the self-energy (Fig.~\ref{fig:FLSE_scaling}), and for the dc resistivity (Fig.~\ref{fig:rho_diff_z}), we find that the optical lineshape is nearly the same in the frequency range $0-0.2$ eV for a range of parameters on a specific iso-$Z$ contour (see Fig.~\ref{fig:optics_scaling}). Again, this demonstrates universality and a weak dependence on the bare model parameters, albeit in a limited frequency range.

Within a Fermi-liquid description, the interacting spectral function should exhibit the same quasiparticle scaling as seen for the self-energy, dc transport, and optical conductivity. Establishing this correspondence provides a direct test of adiabatic continuity in SrVO$_3$ and further examines the extent to which a single quasiparticle coherence scale governs its low-energy electronic properties.

\subsection{Momentum-resolved and summed spectral functions}
\label{sec:spectrum}
\begin{figure}[htbp] 
    \centering
    \includegraphics[scale=0.35]{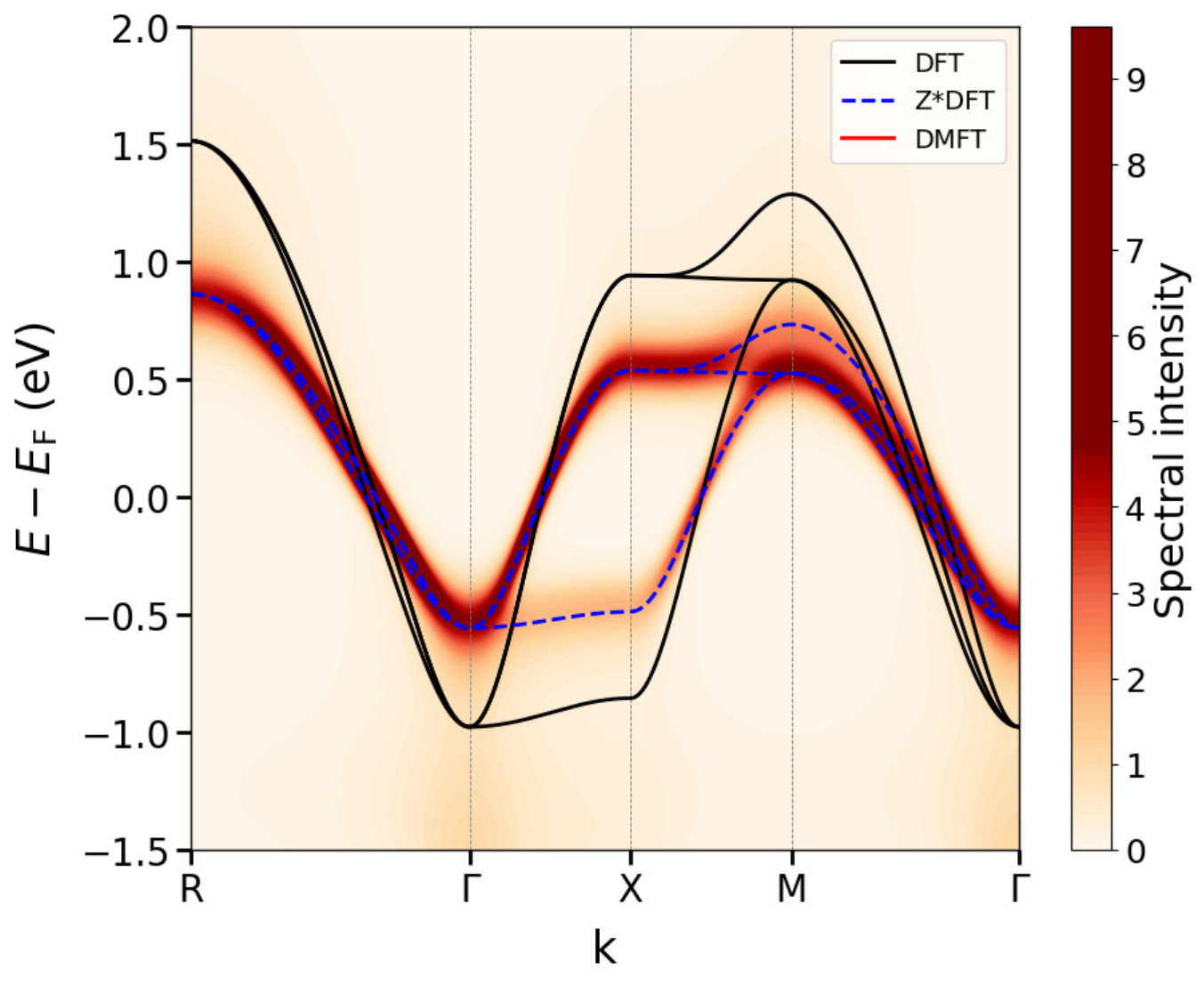}
    \caption{Momentum-resolved electronic structure of SrVO$_3$. The DFT Kohn--Sham bands are shown as thick black lines, while the same bands renormalized by the quasiparticle weight ($Z\approx0.57$) are shown as thick blue lines. The underlying intensity map displays the DMFT momentum-resolved spectral function $A(\mathbf{k},\omega)$ at $T=0$. The low-energy maxima of the spectral function closely follow the renormalized bands, illustrating the adiabatic continuity between the interacting quasiparticle states and the underlying DFT electronic structure.}
    \label{fig:ARPES}
\end{figure}

To this end, we compute the momentum-resolved spectral function $A(\mathbf{k},\omega)$ within the full DFT+DMFT (MO-IPT) framework. The resulting spectral intensity is shown in Fig.~\ref{fig:ARPES}, together with the underlying noninteracting DFT Wannier bands. The low-energy quasiparticle excitations (underlying intensity plot) exhibit a bandwidth reduction relative to the bare Kohn-Sham dispersion, corresponding to a quasiparticle renormalization of $Z\approx0.57$. As illustrated by the blue renormalized bands, the maxima of the interacting spectral function closely track the expected quasiparticle dispersion over the entire coherent energy window, while clearly deviating from the unrenormalized DFT bands. 

\begin{figure}[htbp] 
    \centering
    \includegraphics[scale=0.5]{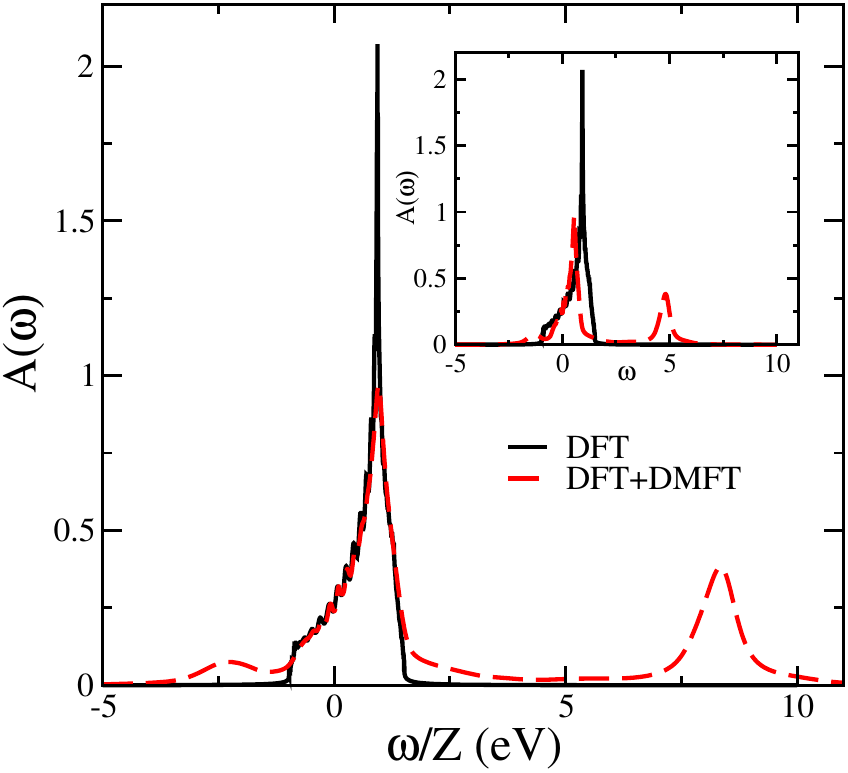}
    \caption{Comparison of the interacting local spectral function of SrVO$_3$, obtained from DFT+DMFT for $U=4.5$~eV and $J=0.65$~eV and rescaled by the quasiparticle weight $Z=0.57$, with the non-interacting DFT spectrum. The collapse of the low-energy spectra onto a common renormalized energy scale illustrates adiabatic continuity and Fermi-liquid quasiparticle scaling. Inset: The same spectra plotted as a function of the bare frequency $\omega$.}
    \label{fig:local_spectra}
\end{figure}

A complementary perspective on this renormalization is provided by a comparison of the local spectral functions obtained from DFT {\it vs.}\ DFT+DMFT, which is shown in Fig.~\ref{fig:local_spectra}. In the main panel, the interacting DFT+DMFT spectrum (red dashed line) is compared directly with the noninteracting density of states  (black solid line) after rescaling the energy axis by the corresponding quasiparticle weight, $Z=0.57$ and $Z=1$ respectively. Remarkably, the low-energy spectral features collapse onto each other, demonstrating that the dominant effect of electronic correlations in the coherent regime is a uniform renormalization of the underlying noninteracting band structure. The inset, plotted as a function of the original unscaled energy $\omega$, highlights the substantial narrowing of the quasiparticle bandwidth induced by correlations, whereas the rescaled representation reveals the near coincidence of the interacting and renormalized noninteracting spectra in the vicinity of the Fermi level. This collapse of spectral features provides a consistency check of theoretically computed DFT and DFT+DMFT spectra through the finding of adiabatic continuity. 

\section{Conclusion}
We have presented a quantitative DFT+DMFT description of the spectral and transport properties of the prototypical moderately correlated metal SrVO$_3$, using a computationally efficient real-frequency multi-orbital iterative perturbation theory (MO-IPT) impurity solver. Over experimentally relevant frequencies and temperatures, the low-energy electronic spectra and transport are governed, to a good approximation, by a single quasiparticle coherence scale: both the spectra and the transport properties are controlled by the same quasiparticle renormalization, which provides a unified description of the spectral function, optical conductivity, and dc transport within a material-specific framework. The interacting spectral function remains adiabatically connected to the underlying non-interacting electronic structure, confirming that the dominant effect of electronic correlations in SrVO$_3$ is a coherent renormalization of the low-energy band structure.

One of the central findings is the weak dependence of the low-energy spectral and transport properties on the bare interaction parameters over the experimentally relevant ranges of $0<\omega<0.2$~eV and $0<T<300$~K: any $(U,J)$ combination yielding a quasiparticle renormalization factor $Z\approx0.5$ (corresponding to an effective mass enhancement of approximately two) gives a quantitatively similar description, since the low-energy physics is largely controlled by this single coherence scale.

The calculated optical conductivity and dc resistivity agree well with experiment. Combining the electron--electron contribution from DFT+DMFT with the electron--phonon contribution from the Bloch--Gr\"uneisen formalism reproduces the observed temperature dependence of ultraclean SrVO$_3$, capturing the Fermi-liquid $T^{2}$ behavior below approximately $25$~K, where electron--electron scattering dominates, and the crossover to phonon-dominated transport at higher temperatures.

A microscopic analysis of the optical conductivity identifies the characteristic low-energy interband feature as arising from transitions between hybridized $t_{2g}$ bands. Decomposing the optical kernel into its phase-space, velocity, occupation, and conductivity contributions isolates the interband transitions responsible for this spectral weight, and calculations in the Wannier representation confirm that suppressing orbital off-diagonal hopping eliminates the feature, establishing interorbital hybridization as its origin. These results corroborate the interpretation of Ahn \textit{et al.}~\cite{GAhn_PRB2022}, with an independent microscopic verification within the present DFT+DMFT framework.

The good agreement between the approximate, real-frequency MO-IPT solver and numerically exact continuous-time quantum Monte Carlo (CTQMC) for the $|\omega| \ll W$ self-energy, obtained at substantially lower computational cost and without the uncertainties of analytic continuation, indicates potential predictive capability of the present framework for the class of correlated multiorbital materials. The work inherits the limitations of the local DMFT approximation, including the neglect of non-local dynamical correlations and vertex corrections in the optical conductivity; extending it to momentum-dependent self-energies via DMFT+$\Sigma_{\mathbf{k}}$~\cite{Sadovskii_PRB2005}, cluster DMFT~\cite{Rohringer_RMP2018}, or related approaches is a natural direction for future work.

\begin{acknowledgments}
The authors acknowledge funding from JNCASR and the computational resources 
provided by the National Supercomputing Mission (\seqsplit{DST/NSM/R\&D\_HPC\_Applications/2021/26}) and JNCASR, India. 
\end{acknowledgments}

\appendix

\section{Nearly universal scaling of self-energy and DC resistivity}
\label{sec:appendix_A}

\begin{figure}[htbp]
    \centering
    \includegraphics[scale=0.5]{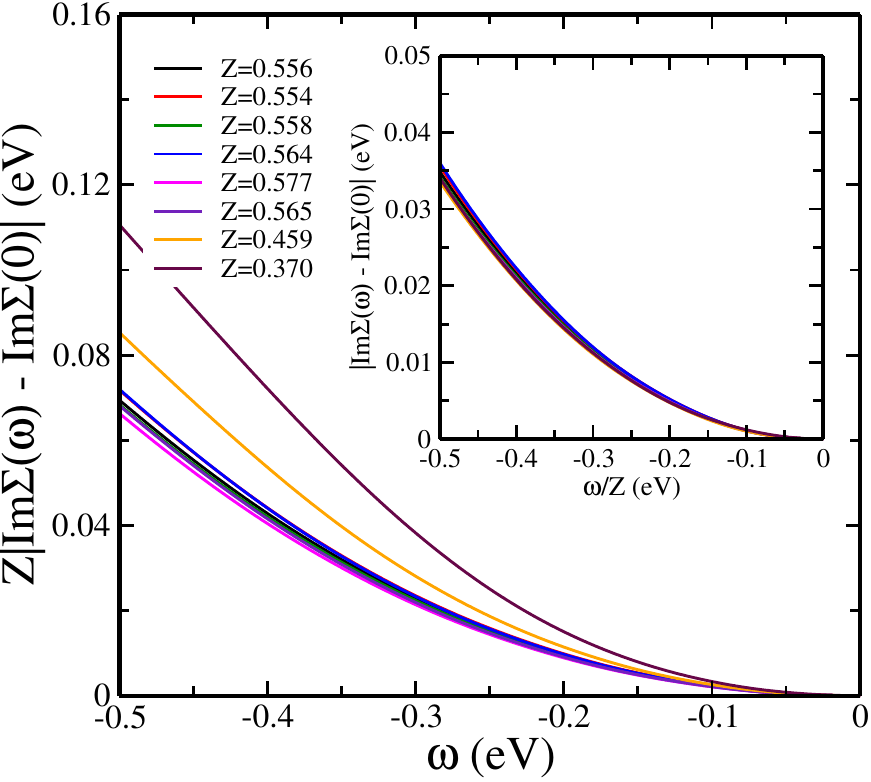}
    \caption{The main panel shows scaled imaginary part of the DMFT self-energy, $Z\,|\mathrm{Im}\,\Sigma(\omega)-\mathrm{Im}\Sigma(0)|$, for representative interaction parameter sets (Table~\ref{tab:UJZ}) at $T=0$~K. Curves sharing a common renormalization $Z\approx0.57$ collapse with little spread, consistent with Fermi-liquid $\omega^2$ scaling up to $\sim-0.5$~eV, while curves at distinct $Z\approx0.46$ and $0.37$ remain visibly separated. Rescaling the frequency axis by each curve's respective $Z$ (inset) collapses all curves onto a single universal form, an effect that emerges upon approaching the strong-coupling limit ($Z\to0$) but is absent near the noninteracting limit ($Z\to1$), signaling that the universality is a genuine correlation effect.}
    \label{fig:FLSE_scaling}
\end{figure}

In this appendix, we present additional results demonstrating the weak dependence of both single-particle and two-particle observables on the bare interaction parameters. Fig.~\ref{fig:FLSE_scaling} extends Fig.~\ref{fig:FLSE} by including several additional interaction parameter sets (see Table~\ref{tab:UJZ}) lying on, or in close proximity to, the iso-$Z$ contour corresponding to $Z\sim 0.57$. The scaled self-energy exhibits a nearly universal collapse over the low-energy window $-0.5 < \omega < 0$~eV. Since one of these parameter sets ($U=4.5$ eV, $J=0.65$ eV) has been shown to provide good agreement with experiment in previous work~\cite{Kugler_PRB2026}, the observed scaling indicates that the low-energy electronic properties depend only weakly on the specific choice of bare interaction parameters, provided the quasiparticle weight is close to its experimental value. By contrast, parameter sets corresponding to distinct renormalizations, $Z\approx0.46$ and $0.37$, yield visibly separated curves, confirming that the collapse is specific to a fixed $Z$ rather than a generic feature of the self-energy. Remarkably, rescaling the frequency axis by the respective $Z$ value of each curve collapses all of these curves, irrespective of their individual $Z$, onto a single universal form. This universal scaling emerges upon approaching the strong-coupling limit ($Z\to0$) but is absent near the noninteracting limit ($Z\to1$), indicating that it is a genuine consequence of strong electronic correlations rather than a trivial feature of the underlying band structure.

\begin{figure}[htbp]
    \centering
    \includegraphics[scale=0.5]{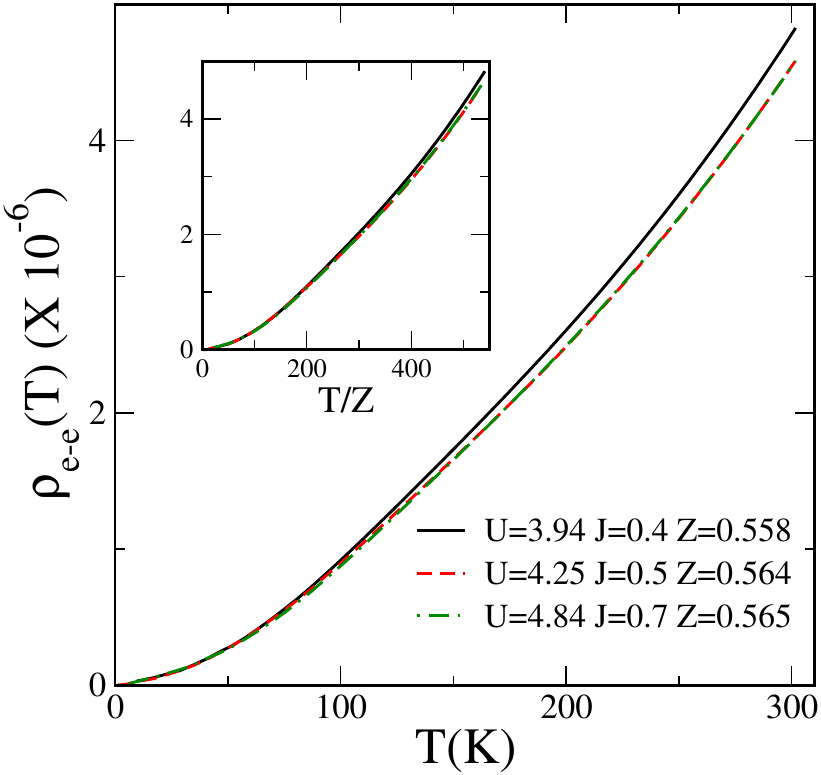}
    \caption{The main panel shows the electron-electron scattering contribution to dc resistivity for three interaction parameter sets $(U,J)$ lying along the iso-Z contour with $Z\approx 0.57$ (see Table~\ref{tab:UJZ}) at $T=0$ K. The near collapse of the curves up to approximately $300$ K illustrates a weak dependence on the interaction parameters. The inset shows the same data with the x-axis rescaled by the quasiparticle weight, where the scaling collapse is far better.}
    \label{fig:rho_diff_z}
\end{figure}

Next, we turn to dc resistivity, which as discussed in the main paper, is a sum of $e-e$ and $e-ph$ contributions. The latter of course does not depend on the bare interaction parameters of the electronic Hamiltonian, while the former could depend on a specific choice of $(U,J)$. However, we show in Fig.~\ref{fig:rho_diff_z} that the $e-e$ contribution to dc resistivity is also nearly universal in the temperature range of $0-300$ K for all $(U,J)$ combinations that yield a single quasiparticle weight. The inset shows that a rescaling of the temperature axis as $T/Z$ yields a much better scaling collapse, since there is a small variability in the $Z$-value of the parameters considered. Thus, any $(U,J)$ combination on the Z=0.57 iso-Z contour yields nearly identical e-e contributions to the dc resistivity over the experimentally relevant temperature range.

\begin{table}[htbp]
\centering
\caption{Representative interaction parameters $(U,J)$ and the corresponding quasiparticle weight $Z$ obtained from the DFT+DMFT (MO-IPT) calculations for SrVO$_3$.}
\label{tab:UJZ}
\begin{tabular*}{\columnwidth}{@{\extracolsep{\fill}}ccc}
\hline\hline
$U$ (eV) & $J$ (eV) & $Z$ \\
\hline
3.39 & 0.20 & 0.556 \\
3.64 & 0.30 & 0.554 \\
3.94 & 0.40 & 0.558 \\
4.25 & 0.50 & 0.564 \\
4.50 & 0.65 & 0.578 \\
4.84 & 0.70 & 0.565 \\
5.13 & 0.80 & 0.577 \\
5.54 & 0.30 & 0.459 \\
8.00 & 0.20 & 0.370 \\
\hline\hline
\end{tabular*}
\end{table}

\section{Microscopic origin of the low-energy interband optical response}
\label{sec:appendix_B}

The optical response of SrVO$_3$ in the noninteracting (Bloch-band) limit is governed by the Kubo expression
\begin{equation}
\sum_{\mathbf{k},mn}
\frac{|v_{mn}(\mathbf{k})|^2}{\Omega}
\bigl[f(E_{n\mathbf{k}})-f(E_{m\mathbf{k}})\bigr]
\delta(E_{m\mathbf{k}}-E_{n\mathbf{k}}-\Omega),
\end{equation}
which follows directly from Eq.~(\ref{eq:sigma_trace}) upon replacing the spectral functions by Dirac delta functions. Here, $v_{mn}(\mathbf{k})$ is the interband velocity matrix element connecting bands $n$ and $m$ at momentum $\mathbf{k}$, $f(E_{n\mathbf{k}})-f(E_{m\mathbf{k}})$ is the difference of Fermi occupation factors that restricts the sum to transitions from an occupied to an unoccupied state, and the delta function enforces energy conservation for a transition of energy $\Omega = E_{m\mathbf{k}}-E_{n\mathbf{k}}$.

To elucidate the microscopic origin of the low-energy optical response in SrVO$_3$, we analyze the distribution of interband transition energies, $\Omega = E_{m\mathbf{k}} - E_{n\mathbf{k}}$, over the full Brillouin zone by constructing a systematic hierarchy of transition histograms that progressively resolves the distinct physical ingredients entering the optical kernel. The full-energy transition density provides a global view of the available phase space for optical excitations and constitutes a discretized representation of the $\delta(E_{m\mathbf{k}}-E_{n\mathbf{k}}-\Omega)$ kernel appearing in the Kubo expression below. It exhibits a strong accumulation of low-energy transitions characteristic of the metallic state, together with a broad continuum arising entirely from interband processes within the $t_{2g}$ manifold.

\begin{figure}[htbp]
    \centering
    \includegraphics[scale=0.1]{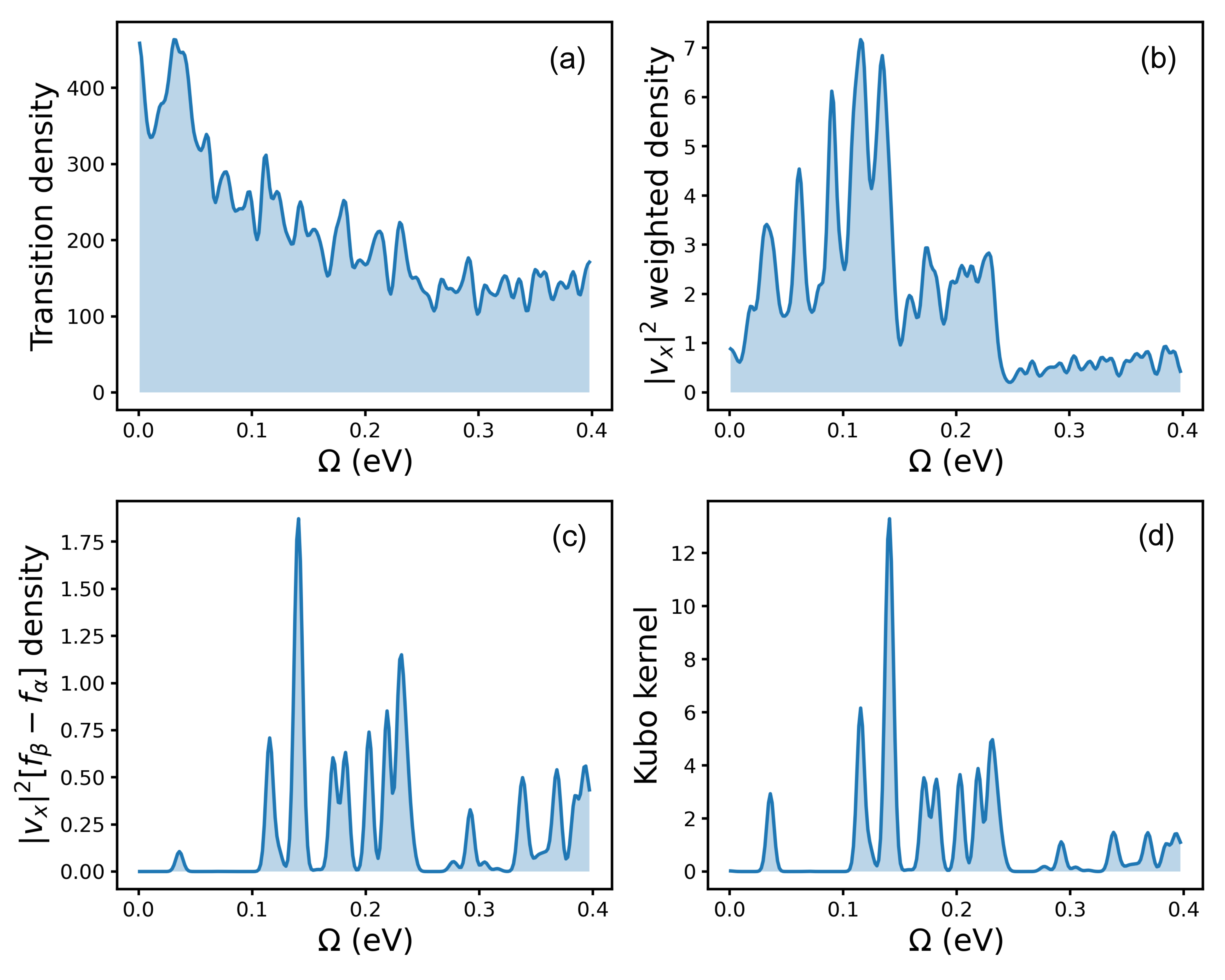}
    \caption{Microscopic decomposition of the low-energy interband optical response in SrVO$_3$ within the Bloch-band representation (Drude contribution removed). (a) Unweighted interband transition density. (b) Velocity-weighted transition density. (c) Velocity- and occupation-weighted transition density. (d) Full non-interacting interband Kubo kernel in the Dirac delta function limit.}
    \label{fig:histograms}
\end{figure}

Restricting the analysis to the optically dominant low-energy sector, $0<\Omega\le0.4$~eV, and to transitions involving at least one quasiparticle state within $1.545$~eV of the Fermi level reveals an unusually large phase-space density of low-energy particle--hole excitations, providing the fundamental $k$-resolved origin of the Drude--interband crossover and explaining the persistence of substantial optical spectral weight well below interband energy scales. Figure~\ref{fig:histograms}(a)--(d) then provides a transparent microscopic decomposition of the optical response through four successive levels of refinement: raw kinematic phase space; optically active transitions weighted by the velocity matrix elements $v_{mn}(\mathbf{k})$; thermodynamically allowed excitations further selected by the Fermi occupation factors $f(E_{n\mathbf{k}})-f(E_{m\mathbf{k}})$; and the full optical integrand, including the explicit $1/\Omega$ factor, prior to any lifetime broadening. The persistence of a pronounced low-energy accumulation of optical weight throughout this hierarchy demonstrates that the low-energy interband response is a robust feature of the underlying electronic structure.

Although the Bloch-band decomposition identifies the optically active transitions, it does not reveal their orbital origin. To establish the underlying microscopic mechanism, we therefore perform a complementary analysis in the Wannier representation by selectively suppressing the orbital off-diagonal hopping terms.

\begin{figure}[htbp]
    \centering
    \includegraphics[scale=0.5]{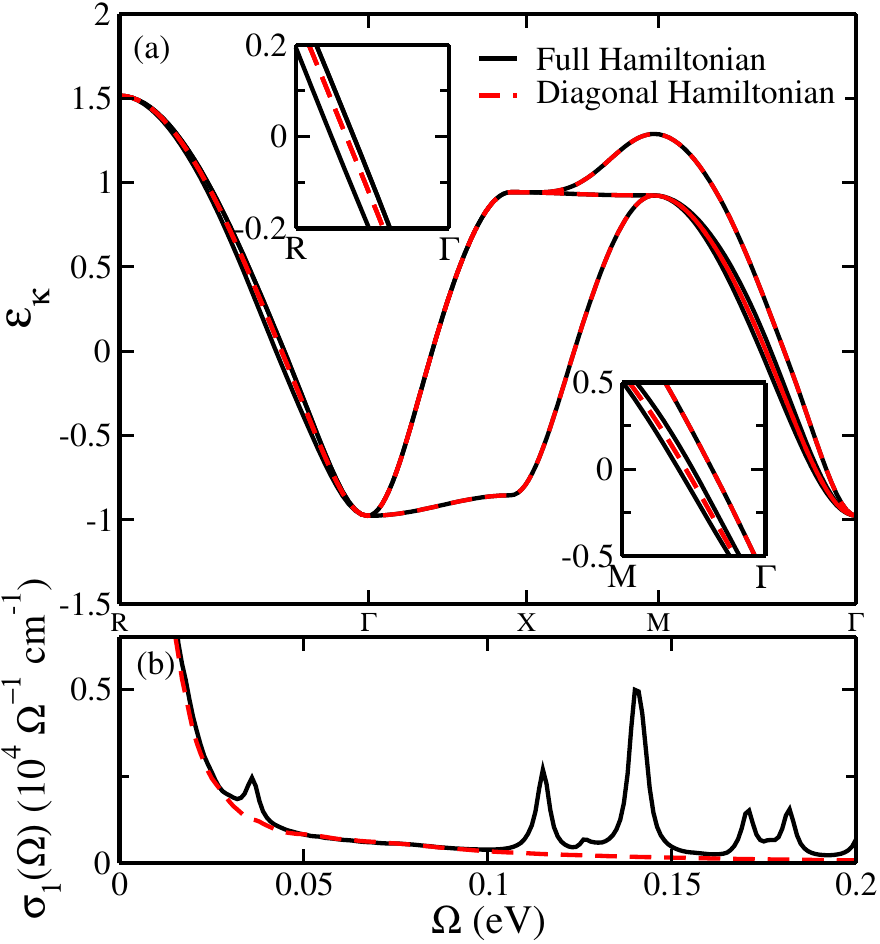}
    \caption{(a)~DFT band structures of SrVO$_3$ obtained from the full Wannier Hamiltonian and from a modified Hamiltonian with the orbital off-diagonal hopping terms suppressed. The insets show enlarged views of the band dispersions along the R--$\Gamma$ and $\Gamma$--M directions. (b)~Corresponding non-interacting optical conductivities computed in the Wannier basis. The pronounced low-energy interband feature present for the full Hamiltonian disappears when the orbital off-diagonal hopping terms are removed, demonstrating that interorbital hybridization is essential for the low-energy optical response.}
    \label{fig:interband}
\end{figure}

Fig.~\ref{fig:interband}(a) compares the band structures obtained from the full Wannier Hamiltonian and from a modified Hamiltonian in which the orbital off-diagonal hopping terms have been removed. Suppressing these terms eliminates the hybridization-induced band splittings responsible for the low-energy interband transitions. Consequently, the corresponding non-interacting optical conductivity, shown in Fig.~\ref{fig:interband}(b), loses the pronounced low-energy interband feature almost entirely. This demonstrates that interorbital hybridization encoded in the Wannier Hamiltonian is the microscopic origin of the low-energy optical response, whereas the orbital-diagonal Hamiltonian alone is insufficient to generate the observed feature. Together with the Bloch-band decomposition of Fig.~\ref{fig:histograms}, these results provide a consistent microscopic picture of the low-energy optical response in SrVO$_3$. Our findings are fully consistent with the mechanism proposed by Ahn \textit{et al.}~\cite{GAhn_PRB2022}, while providing an independent verification within the present DFT+DMFT framework.

We note, however, that the velocity operator employed here is constructed within the Peierls (group-velocity) approximation,
\[
v_{mn}(\mathbf{k})=
\hbar^{-1}
\langle m\mathbf{k}|
\nabla_{\mathbf{k}}H(\mathbf{k})
|n\mathbf{k}\rangle,
\]
which neglects the Berry-connection contribution to the interband matrix elements. The full gauge-covariant velocity operator is given by
\begin{equation}
\hat{v}_{mn}(\mathbf{k})
=
\frac{1}{\hbar}
\langle m\mathbf{k}|
\nabla_{\mathbf{k}}H(\mathbf{k})
|n\mathbf{k}\rangle
+
\frac{i}{\hbar}
(E_{m\mathbf{k}}-E_{n\mathbf{k}})
\mathcal{A}_{mn}(\mathbf{k}),
\label{eq:v_gc}
\end{equation}
where
$\mathcal{A}_{mn}(\mathbf{k})
=
i\langle
u_{m\mathbf{k}}
|
\nabla_{\mathbf{k}}
|
u_{n\mathbf{k}}
\rangle$
is the non-Abelian Berry connection~\cite{Lorenzo2025}. 
In the nearly degenerate and strongly hybridized $t_{2g}$ manifold of SrVO$_3$, this contribution may not be negligible for interband transitions near avoided crossings. Although the Peierls approximation may quantitatively modify the absolute spectral weight of the low-energy feature, its existence and characteristic energy scale are determined by the off-diagonal Hamiltonian structure and therefore remain unaffected. A fully gauge-covariant evaluation of the optical matrix elements, such as the dipole-matrix formalism implemented in the \texttt{woptic} framework~\cite{Assmann2016} and discussed by Wissgott \textit{et al.}~\cite{Wissgott2012}, would provide a useful quantitative refinement of the present analysis.

\section{Nearly universal scaling of optical conductivity}
\label{sec:appendix_C}

\begin{figure}[htbp] 
    \centering
    \includegraphics[scale=0.5]{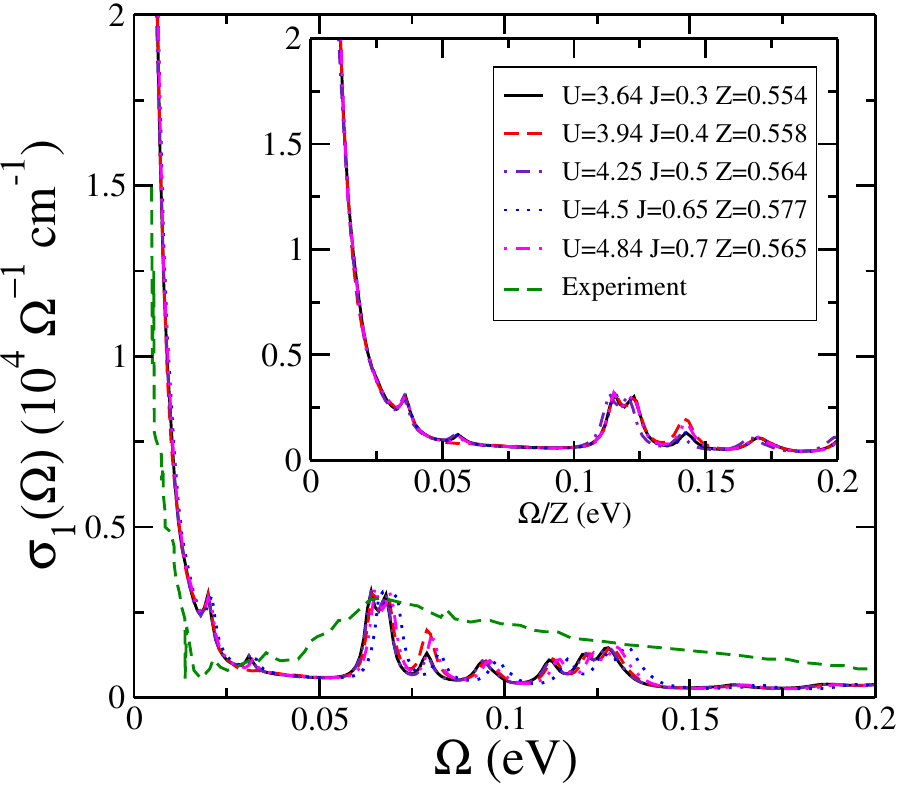}
    \caption{Low-energy optical conductivity of SrVO$_3$ for the representative interaction parameter sets on the $Z\approx0.57$ iso-$Z$ contour (see Table~\ref{tab:UJZ}) at $T=0$~K. Consistent with the universal low-energy self-energy scaling shown in Fig.~\ref{fig:FLSE_scaling}, the low-energy interband feature is nearly independent of the individual values of $U$ and $J$, demonstrating that the optical lineshape upto $\sim 0.2$ eV is governed primarily by the quasiparticle renormalization. The inset replots the spectra as a function of $\omega/Z$, revealing an improved scaling collapse.}
    \label{fig:optics_scaling}
\end{figure}

The optical conductivity, calculated for several $(U,J)$ combinations over the frequency range 0–0.2 eV, likewise exhibits only minor variations (Fig.~\ref{fig:optics_scaling}). This observation is consistent with the universal low-energy scaling of the self-energy and reinforces the conclusion that the low-energy optical response is governed primarily by the quasiparticle renormalization rather than by the individual values of $U$ and $J$. Consequently, any interaction parameter set lying on the $Z=0.57$ iso-$Z$ contour yields a decent agreement with experiments on par with figure~\ref{fig:optics}.

\begin{figure}[htbp] 
    \centering
    \includegraphics[scale=0.5]{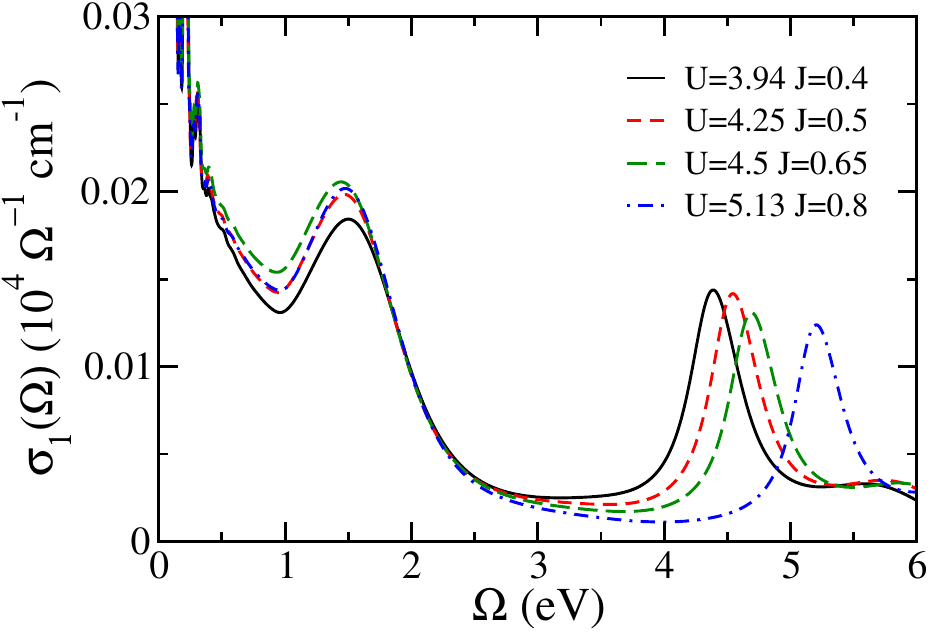}
    \caption{High-energy optical conductivity of SrVO$_3$ for the representative interaction parameter sets on the $Z\approx0.57$ iso-$Z$ contour (see Table~\ref{tab:UJZ}) at $T=0$~K. Unlike the low-energy interband feature, the higher-energy response exhibits a stronger dependence on the interaction parameters. The broad feature near $\sim1.5$ eV is consistent with the conventional mid-infrared (MIR) excitation arising from transitions between the lower Hubbard band and the quasiparticle band, while the broad peak near $\sim5$ eV is associated with transitions between the quasiparticle and upper Hubbard bands.}
    \label{fig:MIR_peaks}
\end{figure}

In contrast, the high-energy optical response exhibits a much stronger dependence on the interaction parameters, as shown in Fig.~\ref{fig:MIR_peaks}. Although all parameter sets lie on the same iso-$Z$ contour and therefore share identical low-energy quasiparticle renormalization, noticeable variations emerge in the position, width, and spectral weight of the broad incoherent features. The peak around $\sim1.5$ eV, commonly identified with the conventional mid-infrared (MIR) excitation involving transitions between the lower Hubbard band and the quasiparticle band, together with the higher-energy excitation near $\sim5$ eV associated with transitions between the quasiparticle and upper Hubbard bands, shifts appreciably with the interaction strength. These results demonstrate that, unlike the universal low-energy optical response governed primarily by the quasiparticle weight, the incoherent high-energy spectrum retains explicit sensitivity to the underlying interaction parameters.

\newpage
\bibliography{references}
\end{document}